\documentstyle[12pt,epsf]{article}

\catcode`\@=11
\def\numberbysection{\@addtoreset{equation}{section}
        \def\theequation{\thesection.\arabic{equation}}}
\def\beq{\begin{equation}}
\def\eeq{\end{equation}}
\def\barr{\begin{eqnarray}}
\def\earr{\end{eqnarray}}
\def\bea{\begin{eqnarray}}
\def\ena{\end{eqnarray}}
\def\winf{W_{1+\infty}\ }
\def\u1{\widehat{U(1)}}
\def\su2{\widehat{SU(2)}_1}
\def\sut{\widehat{SU(3)}_1}
\def\suem{\widehat{SU(m)}_1}
\def\w3{{\cal W}_3}
\def\rep{representation }
\def\reps{representations }

\def\Z{{\bf Z}}
\newcommand{\nl}{\nonumber \\}

\renewcommand{\a}{\alpha}

\numberbysection
\begin{document}
\begin{titlepage}
\begin{center}
\hfill  \quad DFF 318/8/98 \\
\hfill  \quad hep-th/9808179 \\
\vskip .5 in
{\LARGE Hamiltonian Formulation }

{\LARGE of the $\winf$ Minimal Models}

\vskip 0.2in
Andrea CAPPELLI \\
{\em I.N.F.N. and Dipartimento di Fisica, Largo E. Fermi 2,
 I-50125 Firenze, Italy}
\\
\vskip 0.1in
{\em and}
\\
\vskip 0.1in
Guillermo~R.~ZEMBA \\
{\em Centro At\'omico Bariloche and Instituto Balseiro,
C. N. E. A. and Universidad Nacional de Cuyo,
8400 - San Carlos de Bariloche,
R\'{\i}o Negro, Argentina}
\end{center}
\vskip .1 in
\begin{abstract}
The $\winf$ minimal models are conformal field theories 
which can describe the edge excitations of the hierarchical 
plateaus in the quantum Hall effect.
In this paper, these models are described in very explicit terms
by using a bosonic Fock space with constraints, or, equivalently,
with a non-trivial Hamiltonian.
The Fock space is that of the multi-component Abelian
conformal theories, which provide another possible description of the
hierarchical plateaus; in this space, the minimal models are shown to 
correspond to the sub-set of states which satisfy the constraints.
This reduction of degrees of freedom can also be implemented by adding 
a relevant interaction to the Hamiltonian, leading to a 
renormalization-group flow between the two theories.
Next, a physical interpretation of the constraints is obtained
by representing the quantum incompressible Hall fluids as 
generalized Fermi seas.
Finally, the non-Abelian statistics of the quasi-particles in
the $\winf$ minimal models is described by computing
their correlation functions in the Coulomb Gas approach.
\end{abstract}
\vskip 1.cm
\vfill
\hfill August 1998 
\end{titlepage}
\pagenumbering{arabic}
%
\section{Introduction}

The dynamics of the electrons at the hierarchical Hall plateaus 
\cite{prange} has been 
described by several theoretical methods, like the Jain theory of the 
composite fermion \cite{jain}, the effective conformal field theories
\cite{wen} and the Chern-Simons gauge theories \cite{lofra}.
The filling fractions of these plateaus are nicely given by the
Jain formula \cite{jain}:
\beq
\nu={m\over mp\pm 1}\ ,\qquad\qquad m=2,3,\dots,\quad p=2,4,6,\dots\ ,
\label{filfrac}\eeq
which generalizes the Laughlin series \cite{laugh}, $\nu=1,1/3,1/5,\dots$, 
corresponding to $m=1$. 
In this paper we would like to discuss the approach of
conformal field theory (CFT) \cite{cft}; this
describes the low-energy edge excitations, which are
one-dimensional (chiral) waves propagating on the boundary of the sample.
The conformal field theory for the simpler Laughlin plateaus
has been clearly identified as that of the Abelian chiral
boson \cite{wen}, which is also called the chiral Luttinger liquid.
This theory describes the edge waves as well as
the universal properties of the quasi-particle excitations:
in particular, the latter possess fractional electric charge $Q$ and 
fractional statistics $\theta/\pi$, in agreement with the Laughlin 
theory \cite{laugh}.
Some of these properties have been recently measured \cite{expe}
and have been found to agree with the conformal theory.

On the other hand, the edge excitations of the hierarchical plateaus
(\ref{filfrac}) are less understood at present.
Two classes of CFTs  have been proposed: 
the first is the multi-component bosonic theory, which is 
characterized by the Abelian current algebra $\u1^m$ (more precisely,
$\u1\times\suem$) \cite{abe}; these theories generalize the well-established 
one-component theory describing the Laughlin plateaus \cite{wen}.
The other proposed theories are given by the minimal models of 
the $\winf$ algebra \cite{ctz5}. 
According to the Laughlin physical picture \cite{laugh}, 
the electrons form a quantum incompressible fluid at the Hall plateaus.
This fluid can be characterized by the $w_\infty$ symmetry of 
area-preserving diffeomorphisms of the plane \cite{ctz1}\cite{sakita}; 
moreover, the corresponding low-energy edge excitations are described by the
conformal theories with $\winf$ symmetry \cite{cdtz1}, which include 
the specific class of the $\winf$ minimal models \cite{ctz5}.

The two proposed theories are identical for the Laughlin
Hall states; however, they describe the hierarchical
plateaus according to different generalizations of the Laughlin theory. 
They possess the same spectrum of quasi-particles, 
but differ in their specific properties.
So far, the experiments have measured
the quasi-particle charges and the two-point correlation functions,
which are not enough for distinguishing between the two classes of theories.

Let us recall here the main features of the $\winf$ minimal models:
\begin{itemize}
\item 
They exist for the hierarchical plateaus (\ref{filfrac}) only.
\item
They exhibit a smaller number of edge excitations than the
corresponding multi-component Abelian theories.
\item 
They possess a single Abelian charge for the quasi-particles, 
rather than the $m$ charges of the multi-component theories; 
this unique charge is clearly identified with $Q$.
\item
They have neutral quasi-particles which are labelled by the
weights of the $SU(m)$ Lie algebra,
rather than by the $(m-1)$ Abelian charges in the multi-component theories.
As a consequence, the quasi-particles of the minimal theories have
non-Abelian statistics.
\end{itemize}

The $\winf$ minimal models have been introduced in the previous work 
\cite{ctz5}:
they were obtained by the typical algebraic construction of CFT, which
builds their Hilbert space from a consistent collection of 
representations of the symmetry algebra.
Actually, the minimal models are made by the degenerate $\winf$ \reps 
\cite{kac}, which possess Virasoro central charge $c=m=2,3,\dots$, and are
equivalent to the \reps of the $\u1\times {\cal W}_m$ algebra \cite{fz}.
This construction showed that the minimal theories are obtained from 
the multi-component Abelian ones by performing a reduction of
degrees of freedom, which keeps the incompressibility of the electron fluid;
specifically, the reduction of the $\suem$ symmetry down to the
${\cal W}_m$ one.

Afterwards, the exact energy spectrum was numerically computed 
for the Hall states of a system of ten electrons \cite{cmsz}; 
this study clearly indicated that the previous reduction of
degrees of freedom is realized in the spectrum, and
that the $\winf$ minimal models may have a chance to be the correct 
conformal theories of the hierarchical Hall plateaus.
Therefore, it is important to describe this reduction in great detail and
to understand its dynamical origin.

In this paper, we explicitly formulate the $\winf$ minimal models by
quantizing a Hamiltonian of the multi-component bosonic fields.
In Section $2$, this Hamiltonian is shown to be relevant in the 
renormalization-group sense: the corresponding flow
interpolates between the multi-component Abelian and the minimal theories,
and performs the reduction of degrees of freedom for
the latter theory in the infrared fixed point.
This infrared limit can also be obtained by imposing a set of 
constraints on the states of the Abelian Fock space; this 
procedure defines the ${\cal W}_m$ algebra by the so-called
Hamiltonian reduction of $\suem$ \cite{hamred}.
This rather simple description of the $\winf$ minimal models is completely
equivalent to the previous algebraic construction \cite{ctz5}:
moreover, the Hamiltonian formulation may be useful for understanding 
the microscopic dynamics of the reduction and for comparing with other 
approaches and the experimental results.

In Section $3$, we present a simple physical picture for the electron
ground state of the $\winf$ minimal models: this is a 
``minimal'' incompressible Hall fluid, which can be described as a 
chiral Fermi sea, 
as done earlier for the Laughlin fluids \cite{ctz1}\cite{sakita}; 
in this picture, the minimal-model constraints can be interpreted 
as further conditions for incompressibility.
In this Section, we also describe the $\winf$ minimal models in 
terms of spinon excitations, another basis for the $\su2$ states
introduced in Ref.\cite{spinon}.

In Section $4$, we compute the four-point correlation functions of 
the $\winf$ minimal models by using the Coulomb Gas 
method \cite{dofa}. We show that the quasi-particles possess the
non-Abelian statistics; this has a rather simple form, which is 
analogous to that of the [331] double-layer Hall state \cite{nonabe}.
All results are shown in the simplest case of $m=2$ in
(\ref{filfrac}), which is relevant for the plateaus $\nu=2/5,2/9,\dots$
The generalization of this analysis to $m\ge 3$ is described in
the Appendix.

Let us finally stress that this Hamiltonian formulation makes it clear 
that the $\winf$ minimal models are fully consistent theories, although they
are not Rational CFTs \cite{cft}; namely, their partition function is 
not modular invariant \cite{cz} (see the discussion in Section $2$).


\section{The $c=2$ $\winf$ Minimal Models as Reduction of the
Two-component Abelian Conformal Theories}

We first describe the Hilbert spaces of the two
classes of theories by recalling the results of the algebraic 
constructions in Ref.\cite{ctz5} 
(see Section $3.2$). Next, we discuss the equivalent Hamiltonian
description of the minimal models. 


\subsection{The $\u1\times\su2$ Theories}

The two-component Abelian theories are built out of two
chiral bosonic fields $\varphi^{(i)}(R\theta -v_i t)$,
$i=1,2$, which are defined on the edge of the Hall sample \cite{wen}:
here it is chosen to be a disk, parametrized by an angle
$\theta$ and a radius $R$. The conformal field theory is
defined on the space-time cylinder made by the disk 
boundary and time $t$. As usual, the CFT operators are
defined on the complex plane $z=\exp(\tau + i\theta)$, 
where $\tau=it$ is the Euclidean time.
The Fermi velocities $v_i$ as well as the chiralities of the
two bosonic fields are taken positive\footnote{
Namely, we choose the values $\kappa_i=1$ in Section $2$ of 
Ref.\cite{ctz5};
the case of mixed chiralities (some $\kappa_i=-1$) are analogous.}:
this case is relevant for the filling fractions,
\beq
\nu={2\over 2p+1}={2 \over 5}, {2\over 9},\dots 
\label{twojain}\eeq
The edge excitations are described by the chiral currents:
\beq
J^{(i)}= - {1\over 2\pi} {\partial \varphi^{(i)}\over \partial\theta}
={1\over 2\pi}\  \sum_{n=-\infty}^\infty \ 
\a^{(i)}_n {\rm e }^{in\left( \theta -v_i t/R \right)} \ ,\qquad
i=1,2\ ,
\eeq
whose Fourier modes $\alpha_n^{(i)}$ satisfy the
two-component Abelian current algebra,
\beq
\left[\ \a^{(i)}_n \ ,\ \a^{(j)}_m\ \right] =\delta^{ij}\ n \ 
\delta_{n+m,0} \ .
\label{curralg}
\eeq
The corresponding generators of the Virasoro algebra are obtained by
the usual Sugawara construction 
$L^{(i)} = : \left( J^{(i)} \right)^2 :/2$, and read:
\beq
L_n^{(i)} = {1\over 2} \a^{(i)}_0 + \sum_{n=0}^\infty\ 
\a_{-n}^{(i)} \a_n^{(i)} \ ,\qquad\qquad\qquad i=1,2\ .
\label{ln}\eeq
They satisfy
\beq
\left[\ L_n^{(i)} \ ,\ L_m^{(j)} \ \right] = \delta^{ij}
\left\{ (n-m) L_{n+m}^{(i)} + {c\over 12}
n(n^2-1) \delta_{n+m,0} \right\}\ , \qquad c=1\ .
\label{viralg}
\eeq
The generators of conformal transformations are thus given by
$L_n=L^{(1)}_n+L^{(2)}_n$, and obey the same algebra with
$c=2$. The highest weight representations of the Abelian
algebra (\ref{curralg}) are characterized by the
eigenvalues of $J^{(i)}_0$ and $L^{(i)}_0$;
the eigenvalue of $\left(J^{(1)}_0+J^{(2)}_0 \right)$ is
proportional to the quasi-particle charge $Q$ and
$2\left( L^{(1)}_0+L^{(2)}_0 \right)$ is their
fractional statistics $\theta/\pi$.

The Hilbert space of the two-component Abelian conformal theory 
is made of a consistent set of these representations, 
which is complete with respect to the fusion rules: these are the 
selection rules for the composition of
quasi-particle excitations in the theory
(the so-called bootstrap self-consistent conditions) \cite{cft}.
In the Abelian theory, these rules simply require the addition of the
two-component charges of the quasi-particles; as a consequence,
the allowed charge values should fill the points of a two-dimensional lattice.
Each lattice specifies a theory: the adjacency matrix of the lattice 
contains some parameters which are partially determined by the
physical conditions, like the matching of the
filling fraction. Special lattices allow for an extended symmetry 
\cite{abe}: the extension from 
$\u1\times\u1$ to $\u1_{\rm diagonal}\times\su2$
has been chosen because it reproduces the filling fractions
(\ref{twojain}) uniquely (the parameter $p=2,4,\dots$
is related to the compactification radius of the field
$(\varphi^{(1)} +\varphi^{(2)})$ representing the $\u1_{\rm diagonal}$ part).

The corresponding spectrum of quasi-particles was found to be \cite{abe}:
\beq
{\rm I}:\left\{
\begin{array}{l l}
Q & = {2\ell \over 2p+1}\ , \\ 
{1\over 2}{\theta\over \pi} & = {1\over 2p+1} \ell^2 +n^2 \ , \end{array} 
\right. 
\qquad
{\rm II}:\left\{
\begin{array}{l l}
Q & = {2 \over 2p+1}\left( \ell +{1\over 2} \right) \ ,\\ 
{1\over 2}{\theta\over \pi} & = 
{1\over 2p+1} \left( \ell +{1\over 2} \right)^2 +{(2n+1)^2 \over 4}\ ,
\end{array} \right. \quad \ell,n \in {\bf Z}\ .
\label{splitsp}\eeq
In the literature, one often finds a unique formula for the
two cases (I) and (II) above, which can be obtained by substituting
$(\ell,n)\to (n_1,n_2)$, with $(2\ell=n_1+n_2, 2n=n_1-n_2)$
and  $(2\ell+1=n_1+n_2, 2n+1=n_1-n_2)$, respectively 
(see Section $4.2$ of Ref. \cite{ctz5}).
The spectrum (\ref{splitsp}) is clearly factorized in charged and 
neutral excitations, with $\ell$ counting the units of fractional
charge, and $n$ labelling the neutral quasi-particles
of the two-component fluids; the latter are a new feature of the two-component
fluids with respect to the Laughlin ones.
Note that the neutral spectrum is independent of the filling fraction,
i.e. of the value of $p$.

Let us also introduce the current and Virasoro generators in the basis
corresponding to the factorization into charged and neutral sectors
of the spectrum (\ref{splitsp}) (case (I)):
\beq
\begin{array}{llcl}
J = J^{(1)} + J^{(2)} \ , & {J_0\over \sqrt{2p+1}} &
\longrightarrow & Q = {2\ell\over 2p+1}\ ; \\
J^3 = {1\over 2}\left( J^{(1)} - J^{(2)} \right) \ , & J^3_0 & 
\longrightarrow & n \ ;\\
L = L^{(1)} + L^{(2)} = L^Q +L^S \ , & L_0 &
\longrightarrow & {\ell^2 \over 2p+1} + n^2 \ ;\\
L^Q = {1\over 4} \ :\ \left( J\right)^2 \ :\ , \quad 
L^S = \ :\ \left( J^3 \right)^2 \ : \ .& &
\end{array}
\label{coeq}\eeq
(the eigenvalues of case (II) are found for $n\to n+1/2$ and
$\ell\to\ell +1/2$).

Each value in the spectrum (\ref{splitsp}) is the highest weight of
a pair of representations of the Abelian current algebra;
as is well known \cite{cft}, these representations describe 
an infinite tower of edge excitations, generated by the 
bosonic Fock-space operators $\a^{(i)}_n$, $n<0$, $i=1,2$, 
which correspond to Fermionic particle-hole transitions, 
or to their anyonic generalizations \cite{ctz1}\cite{sakita}\cite{cdtz1}.
These excitations are characterized by their angular momentum value,
given by the total Virasoro operator $L_0$; 
moreover, the multiplicities of excitations are counted by the characters
of the 
representations\footnote{
We neglect the charge dependence in the characters, which is
immaterial for the following discussion. See Ref. \cite{cz} for
the complete characters.}:
\beq
\chi^{\u1\times\u1}_{L_0} \equiv
{\rm Tr}_{{\rm Rep}\left(\u1\times\u1\right)}
\left( q^{L_0 -1/12} \right) =
\chi^{\u1}_{\ell^2/(2p+1)} \ \chi^{\u1}_{n^2}\ ,
\eeq
e.g. in the case (I) above; the characters of the Abelian representations
are \cite{cft}:
\beq
\chi_{L_0=h}^{\u1}={q^h \over \eta(q)} \ ,\qquad
\eta(q)=q^{1/24}\ \prod_{k=1}^\infty \left( 1-q^k \right)\ .
\label{abecha}
\eeq
Finally,
the Hamiltonian of the Abelian theory which assigns
a linear spectrum to the edge excitations can be written
in terms of the currents, as follows\footnote{
More general dispersion relations can also be accounted for; see
Ref. \cite{ctz4}.}:
\beq
H= \pi\int_0^{2\pi R} \ dx\ : \left( 
v_1 J^{(1)} J^{(1)} + v_2 J^{(2)} J^{(2)} \right): \ =
{1\over R} \left[ v_1 L_0^{(1)} +v_2 L_0^{(2)} -{1\over 12} \right] \ .
\eeq

Let us now discuss the extended symmetry $\su2$ of the neutral sector 
of the spectrum (\ref{splitsp}).
First, we observe that there are two highest weights with dimension
one, i.e. $n=\pm 1$, which correspond to the additional chiral currents:
\beq
J^\pm = \ : \exp \left( \pm i \sqrt{2}\varphi \right):\ ,
\qquad \varphi = \frac{1}{2}\left( \varphi^{(1)} -\varphi^{(2)} \right)\ .
\label{jpm}\eeq
The two fields $J^\pm$, together with $J^3$ in (\ref{coeq}), form the
$\su2$ current algebra of level $k=1$; their Fourier modes satisfy:
\barr
\left[\ J^a_n\ ,\ J^b_m \ \right] &=& i\epsilon^{abc} J^c_{n+m} +
\frac{k}{2} \delta^{ab} \delta_{n+m,0}\ , \quad k=1 ,\ a,b,c =1,2,3,\nl
\left[\ L^S_n\ ,\ J^a_m \ \right] &=& -m \ J^a_{n+m} \ ;
\label{naca}
\earr
of course, the $J^a_n$ commute with the generators of the charged sector
$\left( L^Q_n, J_m \right)$.
As is well known, there are two highest-weight representations of
the $\su2$ algebra, which are labelled by the ``spin'' $\sigma =0,1/2$.
Their (specialized) characters are \cite{cft}\cite{itz}:
\barr
\chi^{\su2}_{\sigma=0} & = &
{1\over \eta(q)^3}\ \sum_{k\in {\bf Z}}\ \left(6k+1\right)\
q^{(6k+1)^2/12}\ =\ {1\over \eta(q)}
\sum_{k\in {\bf Z}}\ q^{k^2}\ , \nl
\chi^{\su2}_{\sigma=1/2} & = &
{1\over \eta(q)^3}\ \sum_{k\in {\bf Z}}\ \left(6k+2\right)\
q^{(6k+2)^2/12}\ =\ {1\over \eta(q)}
\sum_{k\in {\bf Z}}\ q^{(2k+1)^2/4} \ .
\label{char2}\earr
The second expression for these characters shows that the
two $\su2$ \reps sum up the neutral spectrum of the theory
(\ref{splitsp}) in the two sectors (I) and (II), respectively.
Note also that 
the partition function of the complete theory $\u1\times\su2$ can be
found in Ref. \cite{cz} (for the annulus geometry): 
it includes all the states in the spectrum
(\ref{splitsp}) with multiplicity one and it is invariant under
the modular transformations. Therefore, the $\u1\times\su2$ theory
is a rational CFT \cite{cft}. 

The $\su2$ \reps can be further decomposed into the $c=1$ Virasoro
\reps of weight $h=n^2/4$, $n\in {\bf Z}$.
These are characterized by the property of being 
{\it degenerate}: they contain null-states, i.e.
zero-norm states in their tower of excitations, which should be removed 
from the Hilbert space.
This degeneracy is displayed by the corresponding characters \cite{cft},
\beq
\chi^{\rm Vir}_{L_0^S=n^2/4} = 
{ q^{n^2/4} \left(1-q^{n+1}\right) \over \eta(q)} = 
{ q^{n^2/4} - q^{(n+2)^2/4} \over \eta(q)} \ ;
\label{chiva}\eeq
their numerator contains a negative term which cancels
part of the power expansion of $\eta (q)$, and reduces the 
multiplicities of states with respect to the corresponding
Abelian representation (\ref{abecha}). Actually,
the last part of equation (\ref{chiva}) shows that
each $\u1$ \rep in the neutral spectrum decomposes into
an infinity of Virasoro \reps,
\beq
\chi^{\u1}_{J^3_0=n/2} = \sum_{\ell =0}^\infty\ 
\chi^{\rm Vir}_{L_0^S = (n +2\ell)^2/4} \ .
\label{inver}\eeq
In summary, the decomposition of \reps is given by:
\barr
\{\sigma=0\}_{\su2} &=&
\sum_{k \in {\bf Z}\ {\rm even}}\ \left\{J^3_0={k\over 2} \right\}_{\u1} \ =
\ \sum_{s=0}^\infty \ \left(2s+1\right) 
\left\{L_0^S=s^2 \right\}_{\rm Vir}\ ,
\nl
\left\{\sigma={1\over 2}\right\}_{\su2} &=&
\sum_{k \in {\bf Z}\ {\rm odd}}\ \left\{J^3_0={k\over 2} \right\}_{\u1} =
\sum_{s=1/2,\ s \in {\bf Z}^+ +1/2}^\infty
\ \left(2s+1\right) \left\{L_0^S=s^2 \right\}_{\rm Vir}\ . \nl
&&
\label{deco}\earr
These correspond to the following inclusions of $c=1$ algebras:
\beq
\su2\ \supset\ \u1\ \supset\ {\rm Vir}\ .
\label{inclu}
\eeq

Next, we observe that the decompositions (\ref{deco}) show 
the familiar multiplicities $(2s+1)$ of the $SU(2)$ Lie-algebra \reps
with $s=n/2$ being related to the Virasoro dimension.
Actually, $s$ is a true isospin, because the fusion rules of
the degenerate Virasoro \reps are equivalent to the addition of spins:
$\ \{s+s^\prime\}\oplus\{s+s^\prime-1\}\oplus\dots\oplus\{|s-s^\prime|\}$.
Therefore, the $\su2$ ``spin'' $\sigma$ is only the parity of $s$ (which is
additive modulo two);
moreover, the $\u1$ charge $J^3_0$ is also additive and corresponds to
the projection of the isospin on one axis, i.e. to $m$.

\begin{figure}
\epsfxsize=10cm \centerline{\epsfbox{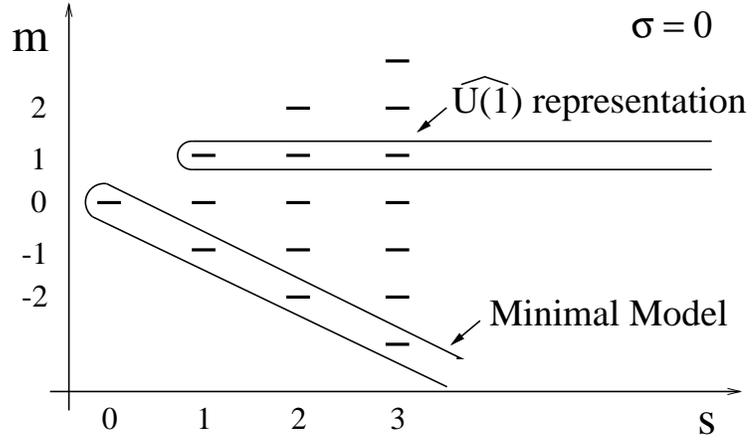}}
\vspace{1cm}
\epsfxsize=10cm \centerline{\epsfbox{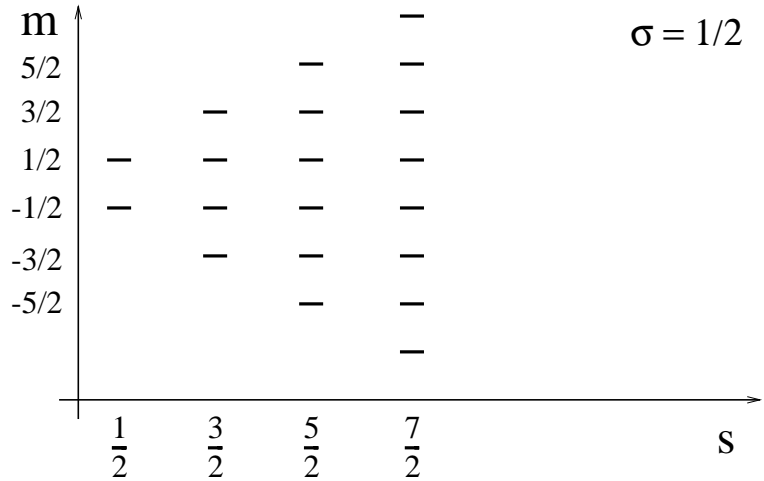}}
\caption{Decomposition of the two $\su2$ \reps in terms of the Virasoro ones:
the horizontal axis is the
total isospin $s$ (Virasoro dimension $L_0^S=s^2$); the vertical axis 
is the isospin component $m$ (eigenvalue of $J^3_0$).}
\label{fig1}
\end{figure}

The generators of this $SU(2)$ algebra inside the $\su2$ \reps
are the zero modes of the currents $\{ J^\pm_0, J^3_0 \}$;
actually, these commute with the $L^S_n$ (see Eq.(\ref{naca})), and thus
act within the $(2s+1)$ Virasoro \reps of same weight.
The decomposition of the two $\su2$ \reps is drawn schematically in Fig.
(\ref{fig1}):
each dash in the $(s, m)$ plane correspond to a Virasoro \rep with
$h=s^2$; the $\u1$ \reps are horizontal arrays
of dashes corresponding to a given $m$ value (see Eq. (\ref{inver})).

\subsection{The $c=2$ $\winf$ Minimal Models}

According to the Laughlin theory \cite{laugh}, 
the ground state at the Hall plateaus
can be described as a two-dimensional incompressible fluid, which is
characterized by constant electron density $\rho_o$
and number $N$. Its low-energy excitations are deformations
of the fluid which have the same area ${\cal A}$:
\beq
N = \int \ d^2x\ \rho \left( {\bf x},t\right)\ \approx\ \rho_o {\cal A}\ .
\label{keyeq}\eeq
These configurations 
can be mapped into each other by area-preserving transformations
of the coordinates of the plane, whose infinitesimal generators 
obey the $w_\infty$ algebra.
Therefore, the (semiclassical) chiral incompressible fluids possess
the $w_\infty$ {\it dynamical} symmetry \cite{ctz1} \cite{sakita}.

The edge excitations are identified with the infinitesimal 
area-preserving deformations of a droplet of fluid; 
their quantization \cite{cdtz1} leads to a conformal field theory with 
the symmetry of the $\winf$ algebra, which is the quantum analog of $w_\infty$.
Note that the $\winf$ algebra contains the $\u1$ and Virasoro algebras.
We refer to our previous works for the description
of the incompressible fluids by the $w_\infty$ transformations \cite{ctz1} and
for the definition of the $\winf$ algebra \cite{ctz4}; here, we recall 
the main properties of the $\winf$ conformal theories.

In Ref. \cite{ctz5}, the $\winf$ theories have been constructed by 
the algebraic method of assembling the \reps of the $\winf$ algebra 
and checking the closure under their fusion rules.
For $c=2$ , the $\winf$ unitary \reps are of
two types: {\it generic} or {\it degenerate} \cite{kac}.
The generic \reps are one-to-one equivalent to those
of the Abelian algebra $\u1\times\u1$ with charges 
$(r^{(1)}, r^{(2)})$ (eigenvalues of $J^{(1)}_0$ and $J^{(2)}_0$)
satisfying $(r^{(1)} - r^{(2)}) \not\in {\bf Z}$.
Clearly, the $\winf$ theories made of generic representations
correspond to the generic two-component Abelian theories.
On the other hand, the degenerate $\winf$ \reps are not equivalent
to the Abelian ones, but are contained into them;
their charges satisfy the condition 
$(r^{(1)} - r^{(2)}) \in {\bf Z}$.
The conformal theories made of the degenerate \reps (only)
have been called the {\it $\winf$ minimal models} \cite{ctz5}.
The main result is that these models are in one-to-one
relation with the hierarchical filling fractions (\ref{filfrac}) 
and yield the same spectrum of charge and fractional statistics
of the $\u1\times\su2$ theories (\ref{splitsp}); 
however, the detailed properties of the two spectra are different,
as anticipated in the Introduction.

The $SU(2)$ ``symmetry'' present in the $c=2$ degenerate 
representations of $\winf$ is a consequence of some properties discussed in 
Ref.\cite{kac}: in general, the $c=m$ degenerate $\winf$ representations 
are equivalent to the representations of the 
$\u1\times{\cal W}_m$ algebra, where ${\cal W}_m$
is the Zamolodchikov-Fateev-Lukyanov algebra at
$c=m-1$ \cite{fz}. The $SU(m)$ Lie algebra is used 
in the construction of the ${\cal W}_m$  algebra, and its
fusion rules are isomorphic to the tensor product of $SU(m)$ representations.
In the case of interest here, $c=2$, we have: 
\beq
{\rm degenerate}\ \winf\ {\rm reps}\ =\ 
\u1\times{\rm Vir}\ \ {\rm reps} \ ,\qquad\qquad (c=2) \ ,
\eeq
where the Abelian algebra describes the charged sector, as usual, 
and ${\rm Virasoro}\sim{\cal W}_2$ accounts for the neutral sector.

After establishing the nature of the degenerate $\winf$
representations, in Ref. \cite{ctz5} the spectrum of the
corresponding minimal models was obtained by checking the fusion
rules: these are again satisfied by a two-dimensional lattice of 
representations.
Their charge and Virasoro weights are still given by the Abelian
spectrum (\ref{splitsp}), but the integer $n$ is restricted to 
$n=0,1,2,\dots$ (a wedge of the lattice), and the meaning of each point in the
lattice is different: in the $\winf$ minimal models,
it represents one $\u1\times{\rm Vir}$ 
representation, while in the $\u1\times\su2$ theories this is a $\u1\times\u1$ 
representation of the same Virasoro weight.

Our next task is to obtain the minimal models as
explicit projections of the $\u1\times\su2$ theories.
Of course, this discussion concerns the neutral sector only.
According to the previous discussion, the minimal model
contains each $h=s^2$ Virasoro \rep with multiplicity one,
while the $\su2$ theories contain $(2s+1)$ copies of it
(see Eq.(\ref{deco})). Therefore, we should choose one 
state per $SU(2)$ multiplet in a consistent way;
this is achieved by imposing a constraint on the states of the form:
\beq
{\cal Q}\ \vert\ {\rm minimal\ state}\ \rangle\ =\ 0\ ,
\label{vinc}
\eeq
for some operator ${\cal Q}$. This constraint is satisfied by 
the states of the $\u1\times\su2$ theory which also
belong to the minimal model; the other states are
projected out.

There are two natural choices for the operator ${\cal Q}$,
namely $J^+_0$ and $J^-_0$; these select the highest
(resp. lowest) state in each $SU(2)$ multiplet.
The previous decomposition of the $\su2$ \reps
into Virasoro ones shows that there are no other choices 
for ${\cal Q}$ in (\ref{vinc}) (see Fig.(\ref{fig1}), 
a more formal derivation will be given in Section $4$).

Therefore, the $\winf$ minimal models can be
defined by the $\u1\times\su2$ theory plus the constraint:
\beq
J^-_0\ \vert\ {\rm minimal\ state}\ \rangle\ =\ 0\ .
\label{minvin}
\eeq
This condition can be completely solved in the basis
$\{ L^S_0, J^3_0 \}$ already discussed at length:
the $SU(2)$ symmetry is clearly killed, because it does not commute 
with the constraint (\ref{minvin}); 
the $U(1)$ quantum number $m$, is also fixed to $m=-s$; 
the remaining good quantum number is the Virasoro weight $h=s^2$.
We can say that ${\cal W}_2$, i.e. Virasoro, is the Casimir sub-algebra
of $\su2$; in general, we have that $\suem \sim SU(m) \times {\cal W}_m$
(see also the Appendix).
Note that the total isospin $s$ continues to compose as before in 
the fusion of two Virasoro representations: namely, in the minimal models,
the composite quasi-particles form $SU(2)$ tensor products,
but there is no $SU(2)$ symmetry.
Some examples of the minimal states satisfying (\ref{minvin})
will be given in Section 3.
The projection (\ref{minvin}) relating the Abelian and minimal
models is a simple case of the general mechanism of 
Hamiltonian reduction of $\widehat{SU(m)}_k$, which is
discussed in the Refs. \cite{hamred}.


\subsection{The Hamiltonian of the $c=2$ Minimal Models}

A first physical interpretation of the constraint (\ref{minvin}) is
that of a strong polarization of the isospin in the $\winf$ minimal
models as opposed to the unpolarized $\su2$ system.
Note, however, that a Zeeman term would give lower
energy to the larger $s$ values, which is not true in the present case.
In order to be more precise, we should introduce the
Hamiltonian which enforces this constraint.
This is given by the following expression ($c=2$):
\beq
H=\frac{1}{R}\left(\ v\ L^Q_0\ +\ v^\prime \ L^S_0\
-\ \frac{1}{12}\ \right)\ +\ \gamma\ J^+_0\ J^-_0\ ,\qquad
\qquad \gamma \in [0,\infty)\ .
\label{hamin}
\eeq
The first term is the standard Hamiltonian for the 
$\u1\times\su2$ theory; the second term is diagonal
in the $(s,m)$ basis, and its eigenvalues are 
$\gamma \left( s(s+1)-m(m-1)\right) $: for $\gamma\to\infty$, it
selects the lowest weight in each $SU(2)$ multiplet, i.e. it
implements the constraint (\ref{minvin}). 

Note that the new term in the Hamiltonian is a non-local interaction:
\beq
J^+_0\ J^-_0\ =\ \oint_{C_t}\ d\theta\ \oint_{C_{t-\epsilon}}\
d\theta^\prime\ {\bf :}{\rm e}^{i\sqrt{2} \varphi(\theta)}{\bf :}\ 
{\bf :}{\rm e}^{-i\sqrt{2} \varphi(\theta^\prime)}{\bf :}\ \ ;
\eeq
actually, this expression cannot be reduced to a single contour integral.
Moreover, this term is relevant in the renormalization-group
sense, because $\gamma$ has the dimension of a mass. 
Therefore, as long as $\gamma$ is switched on, the
infra-red limit of the theory defined by $H$ in (\ref{hamin})
is the $c=2$ $\winf$ minimal model ($\gamma_{eff}=\infty$).
This infra-red fixed point is known exactly because the
$\gamma$ term is diagonal in the considered basis, and,
moreover, it is conformally invariant because 
$[ J^+_0\ J^-_0 , L^S_n]=0$.

Therefore, the Hamiltonian (\ref{hamin}) defines a
renormalization-group trajectory $\gamma\in [0,\infty)$
which interpolates between the two-component Abelian 
theory and the corresponding $\winf$ minimal model.
This renormalization-group flow takes place in the 
same phase of the system, because the two conformal theories
share the same ground-state (which obviously satisfies
the constraint (\ref{minvin})).
Let us remark that this Hamiltonian naturally suggests
the physical relevance of the minimal models: since
the conformal field theories are 
effective low-energy, long-distance descriptions of the
edge excitations, the farthest infra-red
fixed-point is physically relevant; in other words, $\gamma\to\infty$ 
is naturally reached without fine-tuning.

In Ref.\cite{cmsz}, the exact low-lying spectrum has been
computed numerically for ten electrons with repulsive short-range
interactions on the disk geometry.
Two new criteria have been introduced for the 
analysis of the low-lying exact states, which allow for their interpretation
as CFT states and for their identification as either edge or bulk excitations.
A clear pattern is observed:
the edge excitations match the states of the $\winf$ minimal models,
which satisfy (\ref{minvin}); the remaining Abelian states correspond to
bulk excitations.
An energy splitting is found between these two set of states, as suggested by
the Hamiltonian (\ref{hamin}), but is rather weak:
$\gamma \ge O(1/R)$ rather than $\gamma\to\infty$.
In conclusion, the infrared limit cannot be reached in this finite-size
system, but the correct edges excitations are nevertheless
identified with the states of the minimal models described before.

Another virtue of the Hamiltonian (\ref{hamin}) is that it
shows that the $\winf$ minimal models are completely
consistent conformal field theories. 
It had previously been found \cite{cz} that they are not rational CFTs
\cite{cft}, i.e. their partition function cannot by modular invariant\footnote{
See Ref.\cite{kactod} for the proposal of rational extensions
of the $\winf$ models.}.
This is a rather uncommon feature in the literature of CFT, which
remained unexplained: here, we can trace it back to the non-locality
of the $\gamma$ term, which violates one of the hypothesis
for modular invariance \cite{cft}. 
Note also that generalized modular transformations have been 
defined for the $c=1$ (non-minimal) $\winf$ theories \cite{dijk},
for Hamiltonians which contain all possible {\it local} operators $V^i_0$,
$ i=0,1,2,\dots$, in the Cartan subalgebra of $\winf$ \cite{ctz4}:
\beq
H\ =\ \alpha\ J_0\ +\ {v\over R}\ L_0\ + {\beta\over R^2}\ V^2_0 + 
{\delta\over R^3}\ V^3_0 +\ \dots \ ,
\eeq
(here, $J_0\equiv V_0^0$ and $L_0\equiv V^1_0$).
However, the proposed Hamiltonian (\ref{hamin}) does not belong to this
class.

Let us finally remark that the generalization of the constraint
(\ref{minvin}) and of the Hamiltonian (\ref{hamin}) to the 
$\winf$ minimal models with higher central charges $c=3,4,\dots$
is slightly more technical; the case $c=3$ is presented in the Appendix.

\begin{figure}
\epsfxsize=12cm \epsfbox{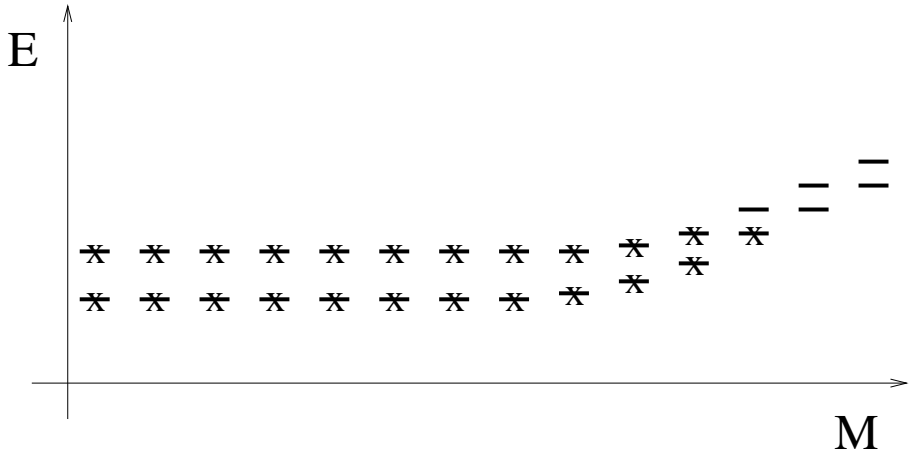} 
\caption{Energy versus momentum plot of 
two filled Landau levels on the disk geometry:
the dashes represent one-particle states, whose energy grows
at large $M$ due to the confining potential;
the crosses represent the electrons.}
\label{fig2}
\end{figure}


\section{Fermionic Realization and the Minimal Incompressible Fluid Picture}
\label{sec3}

In this section, we discuss
the Fermionic Fock space realization of the $(c=2)$ minimal models;  
this leads to the physical picture of the minimal incompressible
Hall fluids as chiral Fermi seas.
Moreover, it allows for a direct comparison with the Jain 
composite-fermion theory \cite{jain}, which has been used 
in the numerical analysis of Ref.\cite{cmsz}.
We first consider the case of two filled Landau levels ($\nu=2$)
(see Fig.(\ref{fig2}));
this is not one of the hierarchical plateaus; thus, this should not
be literally taken as a physical realization of the minimal models.
However, its Hilbert space is isomorphic to those
of the hierarchical states $\nu=2/(2p \pm 1)$, and, in particular,
the neutral sector is identical because it is  
$p$-independent (see Eq.(\ref{splitsp})). 

The edge excitations of two filled Landau levels are described by
two complex chiral (Weyl) fields $\Psi_1$ and $\Psi_2$;
their mode expansion in the Neveu-Schwarz sector is (for $t=0$): 
\bea
\Psi_1 &=& 
\sum_{r \in \Z} {\rm e}^{i\, r\,\theta}\ u_r \ ,\qquad 
{\overline{\Psi}}_1  = \sum_{r \in \Z} 
{\rm e}^{-i\,(r-1)\,\theta}\ u^\dagger_r \ , \nl
\Psi_2  &=& 
\sum_{r \in \Z} {\rm e}^{i\, r\,\theta}\ d_r \ ,\qquad 
{\overline{\Psi}}_2  = \sum_{r \in \Z} 
{\rm e}^{-i\,(r-1)\,\theta}\ d^\dagger_r \ ,
\label{weypl}
\ena
where $u_r$ and $d_r$ denote the Fermionic Fock space operators, which
satisfy $\{ u_k , u^\dagger_l \} = \{ d_k , d^\dagger_l \} =\delta_{k,l}$.
This system defines a $c=2$ conformal field theory with
current algebra $\u1\times\su2$.
The ground state of the system $|\Omega\rangle$ 
is the tensor product of two
Dirac vacua, with corresponding Fermi surfaces (see Fig.(\ref{fig2})). 
One refers to these two vacua as an ``upper'' and ``lower'' components
of $|\Omega\rangle$, which satisfies the conditions:
\bea
u_l |\Omega\rangle &=& d_l |\Omega\rangle = 0 \ ,\quad l > 0~, \nl
u^\dagger_l |\Omega\rangle &=& d^\dagger_l |\Omega\rangle = 0~, 
\quad l \le 0~.
\label{gscond}\ena
In the Landau level picture, $l$ measures the single-particle 
angular momentum with respect to the ground state.

The $\u1\times\su2$ currents of Section $2$ (see Eq.(\ref{coeq}))
can be written in terms of the Fermionic bilinears,
$J^{(i)}\ =\  {\bf :}{\overline{\Psi}}_i  \Psi_i{\bf :}\ , i=1,2 $, 
as follows:
\bea
J\  &=&\ J^{(1)}\ +\ J^{(2)}~~, \qquad 
J^3\ =\ \frac{1}{2} \left( J^{(1)}\ -\ J^{(2)} \right)~~, \nl
J^+\ &=&\ {\overline{\Psi}}_1 \Psi_2  ~~, \qquad \qquad
J^-\ =\ {\overline{\Psi}}_2 \Psi_1  ~~.
\label{curbas}
\ena
Their Fourier modes are given by\footnote{
Of course, their zero modes should be normal-ordered with respect to the vacuum
(\ref{gscond}).}:
\bea
J_n &=& \sum_{k \in \Z} \left(\ u^\dagger_{k-n} u_k\ +\
d^\dagger_{k-n} d_k\ \right)~~, \nl
J^3_n &=& \frac{1}{2}\ \sum_{k \in \Z} \left(\ u^\dagger_{k-n} u_k\ -\
d^\dagger_{k-n} d_k\ \right)~~, \nl
J^+_n &=& \sum_{k \in \Z} u^\dagger_{k-n} d_k~~,\nl
J^-_n &=& \sum_{k \in \Z} d^\dagger_{k-n} u_k~~.
\label{fomcu}
\ena

Using the Sugawara construction, one defines the two $c=1$
stress-energy tensors of Section 2.1, which are associated to the 
charged and neutral sectors, $L^Q = {\bf :} J^2  {\bf :}/4$ 
and $L^S = {\bf :} \left(J^3\right)^2 {\bf :}$, respectively.
Their Virasoro modes can be written in the Fermionic basis as follows:
\bea
L^Q_n &=& \frac{1}{2}\ \sum_{l\in \Z} \left( l - \frac{n+1}{2} \right)
\left( u^\dagger_{l-n}u_l\ +\ d^\dagger_{l-n} d_l \right)\ +\
\frac{1}{2}\ \sum_{l\in \Z}\left( 
\sum_{i\in \Z}\ u^\dagger_{i-n+l}u_i\ \sum_{j\in \Z} d^\dagger_{j-l}d_j
\right)\ ,\nl
L^S_n &=& \frac{1}{2}\ \sum_{l\in \Z} \left( l - \frac{n+1}{2} \right)
\left( u^\dagger_{l-n}u_l\ +\ d^\dagger_{l-n} d_l \right)\ -\
\frac{1}{2}\ \sum_{l\in \Z}\left( 
\sum_{i\in \Z}\ u^\dagger_{i-n+l}u_i\ \sum_{j\in \Z} d^\dagger_{j-l}d_j
\right)\ . \nl
&&
\label{virsq}
\ena

Let us now discuss the excitations of the $\u1\times\su2$ theory
in the Fermionic basis: these 
are labelled by their angular momentum with respect 
to the ground state $\Delta M$, and by the two Abelian
charges $J_0$ and $J^3_0$, which are symmetric and anti-symmetric
with respect to the two layers, respectively.
As explained in Section 2.2, the $c=2$ $\winf$ minimal model
is obtained from this theory by imposing the constraint (\ref{minvin}):
\beq
J^-_0 \vert {\rm \ minimal \ state\ } \rangle =\ 
\sum_{k \in \Z}\ d^{\dagger}_k u_k\ \vert {\rm \ minimal \ state\ }\rangle
=\ 0 \ .
\label{hamr}
\eeq
In this basis, we can interpret the constraint as follows:
the operator $J^-_0$ moves electrons down and holes up 
between the two layers (with a minus sign in the latter case),
without changing their angular momentum value;
it relates the edge excitations in the two layers and actually vanishes
on their symmetric linear combinations.
Therefore, the condition (\ref{hamr}) projects out the edge excitations
which are antisymmetric with respect to the two levels:
this distinguishes the minimal $\winf$ CFTs from the Abelian
ones. Note that the ground state is unique and symmetric, then it satisfies 
the constraint: namely, the two CFTs share the same ground state. 

Let us see some examples of the allowed edge excitations in the minimal 
theory.
There are two Abelian edge excitations at the first excited 
level $\Delta M =1$, which can be written:
\beq
\vert 1; \pm \rangle = \frac{1}{\sqrt{2}} \left( 
d_1^\dagger d_0 \vert \Omega \rangle\ \pm
\ u_1^\dagger u_0 \vert \Omega \rangle \right) .
\label{delta1}\eeq
We find that the symmetric combination 
$\vert 1; + \rangle$ satisfies the
constraint (\ref{hamr}) and the antisymmetric does not. 
At the next level $\Delta M =2$, there are five Abelian edge states:
\barr
\vert\ 2; a \rangle &= &  
d_1^\dagger d_0\ u_1^\dagger u_0 \vert \Omega \rangle\ ,
\nl
\vert\ 2; b\pm \rangle &= &  \frac{1}{\sqrt{2}} \left( 
d_2^\dagger d_0 \vert \Omega \rangle\ \pm
\ u_2^\dagger u_0 \vert \Omega \rangle \right) \ ,
\nl
\vert\ 2; c\pm \rangle &= &  \frac{1}{\sqrt{2}} \left( 
d_1^\dagger d_{-1} \vert \Omega \rangle\ \pm
\ u_1^\dagger u_{-1} \vert \Omega \rangle \right) \ .
\label{deltwo}\earr
Again, the antisymmetric combinations $\vert\ 2; b - \rangle$ and
$ \vert\ 2; c- \rangle$ do not satisfy the constraint (\ref{hamr}). 
Therefore, the $\winf$ minimal models only contains the symmetric 
excitations, as we anticipated.
In general, the number of edge excitations with given $\Delta M$ can be
obtained by expanding the character of the vacuum \rep in powers of $q$
(see Section $2.2$); this counting of states for
the Abelian and minimal theories is reported in Table (\ref{tab1}).
 
\begin{table}
\begin{center}
\begin{tabular}{| c | l | r r r r r r|}
\hline
 & $\Delta M$      & 0 & 1 & 2 & 3 & 4 & 5\ \\ 
\hline 
$c=1$ & One-component Abelian   & 1 & 1 & 2 & 3 & 5 & 7\ \\
\hline
$c=2$ & Minimal Incompressible  & 1 & 1 & 3 & 5 & 10 & 16\ \\
    & Two-component Abelian   & 1 & 2 & 5 & 10 & 20 & 36\ \\
\hline
\end{tabular}
\end{center}
\caption{The number of edge excitations for the Laughlin Hall fluids
$\nu=1/(p+1), p=2,4,\dots$ 
and its quasi-particles (first row) and for the hierarchical fluids 
$\nu=2/(2p+1)$ (second and third rows), 
according to the two relevant conformal field theories.}
\label{tab1}
\end{table}

In Ref. \cite{ctz1}, a physical picture has been developed for the
ground state of the Laughlin plateaus $\nu=1,1/3,1/5,\dots$, 
which is based on the Laughlin incompressible fluid idea.
At the semiclassical level, we said that the deformation
of the incompressible fluids are described by the area-preserving 
transformations obeying the $w_{\infty}$ algebra. 
In the ($\nu=1$) quantum theory, the incompressible fluid can 
be identified as the filled chiral Fermi sea, which occurs in coordinate
space rather than in momentum space (see Fig.(\ref{fig2})); moreover,
the low-lying edge excitations  
are identified with the particle-hole transitions across the Fermi surface. 
The effective quantum theory
for these excitations is given by a single Weyl fermion, which
naturally possesses the $\winf$ quantum symmetry. Actually, the
generators $V^i_n$ of this algebra produce all possible particle-hole 
transitions with $\Delta M=-n$: those with negative $\Delta M$
are not allowed by the Pauli exclusion principle;
thus the ground state is annihilated by all the generators
with positive mode index \cite{ctz1},
\beq
V^i_n\ \vert \Omega \rangle\ =\ 0\ ,\qquad i=0,1,2,\dots, \qquad
n> 0\ .
\label{whwc}
\eeq

These are the $\winf$ highest-weight conditions, which
can be physically interpreted as the incompressibility of the
ground state at the quantum level.
The same incompressibility conditions are 
satisfied by the Laughlin ground states \cite{ctz1}\cite{sakita};
their edge excitations are described by more general 
$c=1$ $\winf$ conformal theories\footnote{
In this case completely equivalent to the $\u1$ chiral-boson theories.}.
Actually, the Hilbert space of the bosonic
($\nu=1/3,1/5,\dots$) and Fermionic ($\nu=1$) theories are
isomorphic and only differ for the quantum numbers of quasi-particles 
\cite{wen}. 

In the case of the hierarchical plateaus, there are, correspondingly,
the present Fermionic basis of two Weyl fermions ($\nu=2$) and
the bosonic basis ($\nu=2/5,\dots$) of the previous Section,
which are again isomorphic, and actually identical for the 
ground state and the neutral sector. 
The $\winf$ highest-weight conditions (\ref{whwc}) on the two-layer Fermi sea 
can be interpreted as before as incompressibility conditions. 
However, the edge excitations possess an additional sector of
up-down antisymmetric particle-hole transitions; these excitations 
are tangential to the 
Fermi surface and cannot be interpreted as deformations of the
incompressible fluid as a whole. In this picture, the
minimal-model constraint (\ref{hamr}) 
can be regarded as an additional incompressibility
condition: it removes the antisymmetric excitations,
which are not necessary for the ``minimal'' incompressible fluids.

Note that the symmetric excitations can be divided
into two classes: those generated by the symmetric current
$J_{-k}, k=1,2,\dots$ in (\ref{fomcu}), which are the same as in
the Laughlin fluids (corresponding to the $\u1_{\rm diagonal}$ part);
in addition, there are those generated by the Virasoro modes 
$L^S_{-k}, k=1,2,\dots$ of the neutral sector, which are specific
of the ($c=2$) minimal incompressible fluid; the first example is
$\vert 2;a \rangle\sim L_{-2}^S \vert \Omega \rangle$ 
in (\ref{deltwo}), which occurs at level $\Delta M =2$
(in agreement with the multiplicities
of the respective theories in Tab.(\ref{tab1})).

This $\nu=2$ description of the minimal models can be pictorially 
extended to the hierarchical states $\nu=2/(2p\pm 1)$. The charged
sector $\u1_{\rm diagonal}$ is bosonized: it remains isomorphic to
the Fermionic one, but the charge eigenvalues change. 
On the other hand, the $p$-independent neutral sector remains 
the same and can still be 
described in the Fermionic basis, provided one pays attention to
distinguishing the two types of symmetric excitations discussed in
the previous paragraph.
This description of the hierarchical plateaus is analog to
the composite-fermion transformation in the Jain theory 
of wave functions \cite{jain} and has been throughoutfully discussed
in Ref.\cite{cmsz}.


\subsection{Spinon Basis}

Another possible basis for the states in the $\su2$ CFT
is given by the spinon excitations \cite{spinon}: the spinon fields,
\beq
\Phi^{\alpha}(z)\ =\ {\bf :}\exp{\left(\ i\ \frac{\alpha}{\sqrt{2}}\ 
\varphi(z)\ \right)}{\bf :}\ , \qquad \alpha = \pm 1\ ,
\eeq
are the $\su2$ primary fields of conformal dimension $h=1/4$ and
isospin $s=1/2$. The exchange of two spinons yields
the fractional statistics $\theta/\pi = 1/2$, which is half of 
that of the fermions; moreover, these fields are not charged, 
since they belong to the neutral sector of $\u1\times\su2$.
They are also called semions, i.e. ``half-fermion'' anyons,
which carry (iso)-spin $1/2$ and no charge.

The Fourier modes of the spinon fields are obtained from the
expansion \cite{bls}:
\beq
\Phi^{\alpha}(z)\ \chi_{\sigma}(0)\ =\ \sum_{n \in\Z}\ 
z^{n+\sigma}\ \Phi^{\alpha}_{-n-\sigma -\frac{1}{4}}\
\chi_{\sigma}(0)\ ,
\label{spimo}
\eeq
where $\chi_{\sigma}(0)$ is a generic state containing an
even (resp. odd) number of spinons for $\sigma=0$ (resp.
$\sigma=1/2$). The $\su2$ Hilbert space can be generated by
applying these modes to the vacuum; this is an alternative 
to the standard basis using the $J^a_n$ modes discussed in
Section $2.1$ (actually, the latter are spinon bilinears).
The spinon fields satisfy generalized (non-local) commutation 
relations which imply relations among the states freely
generated by the multinomials of their modes. These relations
can be casted into the form of a generalized Pauli exclusion
principle: therefore, the modes (\ref{spimo}) build a 
pseudo-Fock space which is neither bosonic ($\alpha^{(i)}_n$)
nor Fermionic ($u_k, d_k$).
This basis has been discussed in the literature in relation 
to the solution of the $SU(2)$ Haldane-Shastry spin chain, whose
spectrum breaks the $\su2$ symmetry down to those of the 
Yangian $Y(sl_2)$ Hopf algebra, which contains $SU(2)$.

This decomposition is rather different from the one relevant for 
the $\winf$ minimal models. Indeed, the corresponding
interaction term of Section $2.3$,
\beq
H^{(2)}\ =\ \gamma\ J^+_0 J^-_0\ ,
\eeq
breaks the $\su2$ symmetry down to Virasoro, $[H^{(2)},L_n]=0$ ;
moreover, the $SU(2)$ symmetry $\{J^{\pm}_0,J^3_0\}$ is
also broken. 
In the case of the Haldane-Shastry chain, the interaction
is given by :
\beq
H_2\ =\ \lambda\ \sum_{k > 0}\ k\ J^a_{-k}J^a_k\ ;
\eeq
this term breaks Virasoro, $[H_2 ,L_n] \neq 0$ for $n \neq 0$,
but keeps scale invariance; moreover, it preserves the
$SU(2)$ symmetry $[H_2,J^a_0]=0$ and the Yangian symmetry
$[H_2,Q^a_1]=0$ generated by the corresponding $Q^a_1$ operators (for their
definition, see Ref. \cite{spinon}\cite{bls}).

Nevertheless, the two breakings of $\su2$ are compatible, because the
interactions in the two problems commute, $[H^{(2)},H_2]=0$.
Actually, we are going to show that the 
constraint 
$J^-_0 \vert {\rm \ minimal \ state\ } \rangle =0$ defining 
the minimal models can be solved in the spinon basis.
First, we remark that the spinon modes 
$\Phi^{\alpha}_{-n-\sigma -\frac{1}{4}}$ have isospin
projection $m= \alpha/2$; this follows from analyzing the 
$\su2$ operator product expansion:
\beq
J^a(z)\ \Phi^{\alpha}(w)\ =\ \left(t^a\right)^{\alpha}_{\beta}\
\frac{\Phi^{\beta}(w)}{(z-w)}\ +\ {\rm regular\ terms}\ ,
\eeq
where $\left(t^a\right)^{\alpha}_{\beta}$ are the Pauli matrices
divided by $2$.
Next, the states of minimal isospin projection $s=-m$, selected by the
constraint $J^-_0 \sim 0$, are simply generated by multilinears of the 
$\Phi^-$ modes, because $m$ is additive. Therefore, the
fully polarized $N$-spinon states (see Ref. \cite{bls}), 
\beq
\Phi^-_{-\frac{(2N-1)}{4}-n_N}\dots\Phi^-_{-\frac{5}{4}-n_3}\
\Phi^-_{-\frac{3}{4}-n_2}\ \Phi^-_{-\frac{1}{4}-n_1}\ \vert
\Omega\rangle\ , 
\label{spba}
\eeq
are explicit solutions of the constraint
$J^-_0 \sim 0$. The Virasoro weight of these states is given by:
\beq
h\ =\ \frac{N^2}{4}\ +\ \sum_{i=1}^N\ n_i\ ;
\label{hdim}
\eeq
clearly, an even (resp. odd) number of spinons gives a 
state in the $\sigma=0$ (resp. $\sigma=1/2$) $\su2$ representation
(see Fig.(\ref{fig1})).

It remains to be established a basis of independent states among 
all possible $N_{th}$-plet $\{n_i \}$ in (\ref{spba}). This is
a rather difficult task, because the ``commutation relations''
among the modes contain infinite terms; therefore, we shall use 
an indirect argument. We start from the basis of independent 
states made by $N^+$ up and $N^-$ down spinons, which has been
introduced in Ref. \cite{bls} (Basis II):
\barr
& &\Phi^-_{-\frac{2(N^+ + N^-)-1}{4}-n^-_{N^-}}\cdots
\Phi^-_{-\frac{2N^+ + 1}{4}-n^-_1 }\
\Phi^+_{-\frac{2N^+ - 1}{4}-n^+_{N^+} }\ \cdots
\Phi^+_{-\frac{1}{4}-n^+_1}\ \vert
\Omega\rangle\ , \nl
& &\ \ n^+_{N^+}\ge\dots\ge n^+_2 \ge n^+_1\ge 0\ ,
\qquad n^-_{N^-}\ge\dots\ge n^-_2 \ge n^-_1\ge 0\ .
\label{bastwo}
\earr
The Virasoro dimension is again given by (\ref{hdim}) with
$N=N^+ + N^- $ and $\{n_i\}=\{n^+_i ,n^-_j \}$. It has been shown 
\cite{bls}, that this basis reproduces the $\su2$ representations 
$\sigma=0$ for $N^+ + N^-$ even (resp., $\sigma=1/2$ for $N^+ + N^-$ odd);
this can be checked by computing the corresponding characters,
using the following identity for the sum of partitions:
\beq
\sum_{k_1\geq \dots k_N\geq 0}\ q^{\sum_{i=1}^{N}\ k_i}\ =\
\prod_{\ell =1}^N\ \sum_{m_{\ell}=0}^{\infty}\ q^{\ell\ m_{\ell}}\ =\
\frac{1}{\prod_{k=1}^N \left( 1-q^k \right)}\ \equiv\ 
\frac{1}{\left( q\right)_N}\ .
\eeq
One obtains:
\beq
\chi^{\su2}_{\sigma=0}\ =\ 
\sum_{N^+ ,N^- =0,\ N^+ + N^- {\rm even}}^{\infty}\
\frac{q^{(N^+ + N^-)^2/4}}
{\left( q\right)_{N^+}\  \left( q\right)_{N^-}} \ ,
\eeq
and similarly the sum extended to $(N^+ + N^-)$ odd for 
$\sigma =1/2$. These are actually equivalent expressions for 
the characters of the $\su2$ \reps (\ref{char2}), owing to the
identities discussed in Ref. \cite{melzer}.

Next, we remark that the states (\ref{bastwo}) have isospin
projection $m=(N^+ -N^-)/2$ and isospin $0\le s \le (N^+ + N^-)/2$.
Consider the subspace of states with $J^3_0=m=0$,
i.e. $N^+ = N^-$: their sum reproduces the $\u1$ vacuum 
representation, as is clear from Fig. (\ref{fig1}).
One can check the corresponding character:
\beq
\chi_{h=0}^{\u1}\ =\ \sum_{N^+ =0}^{\infty}\ \frac{q^{(N^+)^2}}
{\left( \left(q\right)_{N^+} \right)^2}\ =\
\frac{1}{\eta(q)} \ .
\eeq
Similarly, the $m=-1/2$ subspace, $N^- = N^+ +1$, yields:
\beq
\chi_{h=1/4}^{\u1}\ =\ \sum_{N^+ =0}^{\infty}\ \frac{q^{(2N^+ +1)^2/4}}
{\left(q\right)_{N^+}\left(q\right)_{N^+ +1} }\ =\
\frac{q^{1/4}}{\eta(q)} \ .
\eeq
{}From Fig.(\ref{fig1}) and the discussion in Section $2.1$, it follows
that the $m=0$ subspace is isomorphic to the Hilbert space 
of the minimal model, in the sector $\sigma=0$ (the $m=-1/2$
one matches the $\sigma=1/2$ sector, respectively). 
The mapping can be obtained by multiple application of $J^-_0$, as it follows: 
given that $s\leq N^+$ for the states with $N^+ = N^-$ (\ref{bastwo}),
the action of $\left( J^-_0 \right)^{N^+}$ yields the states
of minimal projection $m=-s=-N^+$ which are required for the minimal models;
otherwise, it vanishes.
Moreover, this action replaces all the $N^+$ modes $\Phi^+_{\beta}$ with 
$\Phi^-_{\beta}$ in Eq. (\ref{bastwo}). Therefore, the basis 
(\ref{bastwo}) is mapped into the one of fully
polarized spinons (\ref{spba}), which satisfy the constraint 
$J^-_0 \sim 0$.

In conclusion, the (neutral part of) Hilbert space of the 
$c=2$ $\winf$ minimal models can be described
by the fully-polarized spinon states (\ref{spba}) with the 
following choices of $\{ n_i \}$:
\barr
\sigma =0\ : \ \ &N&\ {\rm even}\ ,\ 
\left\{ { {n_N \geq n_{N-1}\geq\dots\geq n_{\frac{N}{2} +1} \geq 0}\ ,
\atop {n_{\frac{N}{2}-1}\geq n_{\frac{N}{2}-2}\geq\dots\geq n_1 
\geq 0} \ ;} \right. \nl
\sigma = \frac{1}{2}\ :\ \ &N&\ {\rm odd}\ ,\ \  
\left\{ { {n_N\geq n_{N-1}\geq\dots\geq n_{\frac{N+1}{2}} \geq 0}\ , 
\atop {n_{\frac{N-1}{2}} \geq n_{\frac{N-1}{2}-1}\geq\dots\geq n_1 
\geq 0} \ .} \right.
\earr
Note that this basis is larger than the one introduced in Ref. \cite{bls}
(Basis I) for building the Yangian highest-weight states (as it should).
Let us finally remark that the physical interpretation of this
basis for the ($c=2$) minimal models remains to be understood,
possibly by extending the picture of the Fermi sea 
discussed in the previous paragraph.   
 

\section{Correlation Functions and Non-Abelian Statistics}

In this Section, we derive the correlation functions of the
quasi-particles in the $c=2$ $\winf$ minimal models, i.e. the 
$\u1\times{\rm Vir}$ CFTs. These correlators factorize into
charged and neutral parts; the charged $N$-point functions are
given by the well-known vertex operator expectation values \cite{cft}:
\barr
&&\langle\ \Omega\ \vert\ V_{Q_1}(z_1)\dots V_{Q_N}(z_N)\ \vert\ \Omega\
\rangle\ =\ \prod_{i<j}^{N}\ \left( z_i - z_j 
\right)^{Q_i Q_j \left(p+\frac{1}{2}\right) }\ , \nl
&&V_Q (z)\ =\ {\bf :} \exp{\left(i\ Q\ \sqrt{2p+1}\ 
\left( \varphi^{(1)}(z)\ +\ \varphi^{(2)}(z) \right)
\right)} {\bf :}\ ,
\label{verco}
\earr
where the $Q_i$ are given by the spectrum (\ref{splitsp}).

The correlators of the neutral part can be obtained by taking the
$c\to 1$ limit of the Dotsenko-Fateev Coulomb Gas construction \cite{dofa},
which applies to the $c \leq 1$ Virasoro minimal models.
In this approach, the $N$-point functions are written in terms of
the $\u1$ vertex operators plus screening charges $Q_{\pm}$, which
project the Virasoro null states out of the $c=1$ bosonic Fock space.
The properties of the screening charges have been investigated by 
Felder and other authors \cite{felder}: we first recall some relevant
formulae of these works, then we properly take the $c \to 1$ limit,
and compare the result with the formulation of Section $2$.
Next, we compute some examples for the correlators and discuss the
non-Abelian statistics.

Let us start from the $\u1$ theory of the neutral bosonic field 
$\varphi = \left( \varphi^{(1)} - \varphi^{(2)} \right) /2$
(Section $2$); the screening charges are defined by: 
\beq
Q_{\pm}\ =\ \oint\ dz\ {\bf :}\ {\rm e}^{i\sqrt{2} \alpha_{\pm} \varphi(z)}\ 
{\bf :}\ ,
\eeq
with
\beq
\alpha_{+} =\sqrt{\frac{p'}{p}}\ ,\qquad 
\alpha_{-} =-\sqrt{\frac{p}{p'}}\ ,\qquad 
c =c(p,p') = 1 - 6 \frac{(p-p')^2}{pp'} \leq 1\ ,\qquad
p,p' >0\ .
\label{vimi}
\eeq
These operators have vanishing conformal dimension in the $c(p,p')$
CFTs, and satisfy $[Q_{\pm} , L_n ]=0$. 
Their action in the bosonic Fock space is to relate states
of same conformal dimension which belong to different $\u1$
representations:
\beq
F_{\a_{m',m}}\ {\buildrel Q_+ \over \longrightarrow }\
F_{\a_{m',m-2}}\ , \qquad
F_{\a_{m',m}}\ {\buildrel Q_- \over \longrightarrow }\ F_{\a_{m'-2,m}}\ ,
\eeq
where we denote with $F_{\a_{m',m}}$ the representation with ``charge''
$J^3_0 = \a_{m',m}$. Its relevant values are:
\barr
\a_{m',m} &=& \a_0 - m \a_+ - m' \a_- \ ,\qquad
2 \a_0 = \a_+ + \a_-\ , \nl
h_{m',m} &=& \a_{m',m} \left(\a_{m',m} -2\a_0\right) =
\frac{(m' p-mp')^2-(p-p')^2}{4pp'}\ ;
\label{kactab}\earr
actually, they reproduce the dimensions $h_{m',m}$ 
of the $c<1$ degenerate Virasoro representations (the Kac table) \cite{cft}.

The screening charges are useful because they can map $\u1$ 
highest-weight states into Virasoro null states, and thus
describe their structure. The two lowest null states in the 
Virasoro representation $h=h_{m',m}$ have dimensions \cite{cft}:
\beq
h_{m',m}\ +\ m'm\quad ,\quad h_{m',m}\ +\ (p'-m')(p-m)\ .
\label{nullvec}\eeq
We are interested in the $c=1$ degenerate Virasoro representations,
which can be obtained by the controlled limit
$c(p,p')\to 1$ within the Virasoro minimal models.
This limit can be achieved by letting,
\beq
p\ ,\ p'\ \to\infty\ ,\qquad \frac{p}{p'}\ \to 1\ ,
\eeq
which implies:
\beq
\a_{\pm}\to 1\ ,\ \a_0\to 0\ ,\ \a_{m',m}\to \frac{(m'-m)}{2}\ 
\equiv \frac{n}{2}\ ,\quad h_{m',m} \to \frac{n^2}{4}\ .
\eeq
The screening charges go into the $SU(2)$ operators, $Q_{\pm} \to
J^{\pm}_0$, and the $c<1$ degenerate representations are mapped
into the $c=1$ ones, $h = n^2/4$, with infinite
degeneracy.
It can be shown that all the $c<1$ null states but the lowest one
decouple (their dimensions go to infinity). 
In Felder's notation \cite{felder}, $w_0$ denotes the 
remaining null state in the $c=1$ representation with $h=n^2/4$;
this state has the dimension
$h=n^2/4 +n+1 = (n+2)^2/4$, e.g. $m=n+1$, $m'=1$ in (\ref{nullvec}); 
moreover, it can be obtained from the
$\u1$ highest-weight $v_0$ with charge $J^3_0 = -(n+2)/2$ by the 
action of $Q_+ \equiv J^+_0$:
\beq
\left( w_0 \right)_{J^3_0= -n/2}\ =\ J^+_0\ 
\left( v_0 \right)_{J^3_0= -(n-2)/2}\ .
\label{felc}
\eeq
This result matches the description of Section $2$: from Fig.(\ref{fig1})
and the characters (\ref{char2}), it is apparent that the $h=n^2/4$
Virasoro \rep with $s= -m=n/2$ (the $SU(2)$ lowest-weight state) 
can be obtained from the $\u1$ \rep
with $m=-n/2$ (horizontal line in Fig.(\ref{fig1})) by subtracting
the $\u1$ \rep of the null-vector with $m=-n/2 -1$ (adjacent horizontal line). 

In conclusion, the $c=1$ Virasoro theory is described in the Felder
approach \cite{felder} as the $\u1$ Fock space with all the 
null-vectors removed;
according to Eq.(\ref{felc}), this condition can be written: 
\beq
\left\vert\ {\rm physical\ state}\ 
\left( J^3_0 =-m\right) \right\rangle\ \neq\ 
J^+_0\ \left\vert\ {\rm any\ state}\ 
\left( J^3_0 =-m-1\right) \right\rangle\  
\qquad (m \geq 0)\ .
\eeq
This is clearly equivalent to the constraint 
$J^-_0 \vert\ {\rm physical\ state}\ \rangle =0$ introduced in
Section $2$;
therefore, the Felder (Coulomb Gas) approach yields another 
description of the Hilbert space of the $\winf$ minimal models, 
which is equivalent to the previous ones.


\subsection{Examples of Correlators}

In the Dotsenko-Fateev Coulomb Gas approach \cite{dofa},\cite{cft}, 
the $N$-point function of Virasoro primary fields $\phi_h$ is obtained by
the vertex operators with appropriate insertion of a number of
screening charges; for example, the four-point functions have the form:
\barr
\label{fpf}
&&\langle\ \phi_{s_1^2}(z_1)\phi_{s_2^2}(z_2) \phi_{s_3^2}(z_3)
\phi_{s_4^2}(z_4)\ \rangle_{\rm Vir}\ =\\
&&\langle\ \Omega\ \vert\ \left( \prod_{i=1}^{N_+} Q_+ \right)
{\bf :}{\rm e}^{-im_1 \sqrt{2} \varphi(z_1)}{\rm :}
{\bf :}{\rm e}^{-im_2 \sqrt{2} \varphi(z_2)}{\rm :}
{\bf :}{\rm e}^{-im_3 \sqrt{2} \varphi(z_3)}{\rm :}
{\bf :}{\rm e}^{-im_4 \sqrt{2} \varphi(z_4)}{\rm :}
\ \vert\ \Omega\ \rangle\ ;\nonumber
\label{fourp}\earr
in this expression, the number of screening charges is determined
by the condition of charge neutrality:
\beq
J^3_0 = N_+ - m_1 - m_2 - m_3 - m_4 = 2\a_0 = 0\ ,\qquad\quad
(c=1)\ ,
\label{tocha}
\eeq
with $m_i^2=s_i$, $i=1,2,3,4$.
In the $c<1$ minimal models, both screening charges 
$Q_+$ and $Q_-$ (see Eq.(\ref{vimi})) must be used in 
the correlator (\ref{fpf}), in order to satisfy the charge
neutrality condition (see Eq.(\ref{kactab}));
in the $c=1$ case, there are several possibilities, because the
two $Q_{\pm}$ charges are equivalent by the $m \to -m$ symmetry
(absent for $c<1$); moreover, in some correlators the charge
neutrality condition can also be satisfied without screening 
charges, provided that certain permutation rules are employed
(see later).

Let us first consider the example of the three-point function of
the fields $s_1=s_2=k/2$ and $s_3=n$:
\beq
\langle\ \Omega\ \vert\ \left( \prod_{i=1}^{k-n} Q_+ \right)
{\bf :}{\rm e}^{-i{k \over\sqrt{2}} \varphi(z_1)}{\rm :}
{\bf :}{\rm e}^{-i{k\over \sqrt{2}} \varphi(z_2)}{\rm :}
{\bf :}{\rm e}^{in\sqrt{2} \varphi(z_3)}{\rm :}
\ \vert\ \Omega\ \rangle\ .
\eeq
Using the reflection symmetry $m_i \to \pm m_i$ for each field,
one can choose the simplest case of minimal number of screening
charges.
Moreover, we can take $z_1=0,z_2=1,z_3\to\infty$ by using the
$SL(2,{\bf C})$ invariance of the ground state.
A number of properties can be derived from this correlator: as
an example, we check the $SU(2)$ fusion rules 
$\{k/2\}\times\{k/2\}=\{0\}+\{1\}+\dots+\{k\}$; these imply that
the correlator should not vanish for $n=0,1,\dots,k$. For
$n=k$, it obviously does not; for $n=k-1 \geq 0$, we find the
expression: 
\beq
\oint_{C_{u_0}}\ du\ u^{-k}\ (1-u)^{-k}\ ,
\label{tpc}
\eeq
where the contour $ C_{u_0}$ has to be taken around any of the
singularities $u_0=0,1,\infty$. This integral is found to vanish
for $C_{\infty}$ and to take opposite non-vanishing values for
$C_0$ and $C_1$. Next,
for $n=k+1$, the charge neutrality condition requires the use of 
one $Q_+$, leading to the same expression (\ref{tpc}) with
$k \to -k$; this vanishes for any choice of $ C_{u_0}$ and verifies
again the fusion rules. Other cases for $n$ can be worked out along 
the same lines \cite{dofa}.

The non-Abelian statistics of quasi-particles manifests itself in the
four (and higher) point functions, as we shall now discuss. 
In the Coulomb Gas form of
the correlator (\ref{fpf}), the choices of contour $ C_{u_0}$ yield
several independent expressions, which are called the
conformal block ${\cal F}_i$ \cite{cft}: each one corresponds to 
an intermediate state with a given isospin value, obtained
by the fusion of the isospins involved.
For example, the correlation of four electron excitations
($Q=1$ and $s_i=1/2$ in (\ref{fourp})) can be written:
\barr
\label{cbc}
G(z_i)&=& \langle\ \Omega\ \vert\ Q_+ 
{\bf :}{\rm e}^{-{i \over\sqrt{2}} \varphi(z_1)}{\rm :}
{\bf :}{\rm e}^{-{i\over\sqrt{2}} \varphi(z_2)}{\rm :}
{\bf :}{\rm e}^{-{i\over\sqrt{2}} \varphi(z_3)}{\rm :}
{\bf :}{\rm e}^{{i\over\sqrt{2}} \varphi(z_4)}{\rm :}
\ \vert\ \Omega\ \rangle\ \nl
&=& { (\eta(1-\eta))^{1/2}\over (z_{13} z_{24})^{1/2} }\
\oint_{C_i}\ du\ u^{-1}\ (u-1)^{-1}\ (u-\eta)^{-1}\\
&=&{1 \over (z_{13} z_{24})^{1/2} }\ 
\left[ \ A_1\ \left(\frac{1-\eta}{\eta}\right)^{1/2}+
A_2\ \left(\frac{\eta}{1-\eta}\right)^{1/2}+
A_3\ \left(\frac{1}{\eta(1-\eta)}\right)^{1/2}\ \right]\ .\nonumber
\earr
In this expression, we have used the $SL(2,{\bf C})$ invariance
to map the four points $(z_1,z_2,z_3,z_4) \to (0,\eta ,1,\infty )$,
with $\eta=z_{12} z_{34}/(z_{13} z_{24})$, and $z_{ij} \equiv (z_i - z_j)$.
The correlator (\ref{cbc}) contains two independent blocks for the
isospin channels $s=0,1$, which are added with coefficients $A_1$ and
$A_2$; the third term in
the last line of (\ref{cbc})
is actually a linear combination of the first two.
Note also that the full correlator also contains the factor 
$\prod_{i<j}\ z_{ij}^{p+1/2}$ arising from the charged part.

The presence of two independent terms in the correlator (\ref{cbc})
should be interpreted as a ``dynamical'' degeneracy of the
four-electron state. Under the exchange of two electrons, those 
two terms transform linearly into themselves. This is 
the notion of non-Abelian statistics: the exchange operation does
not simply map each state into itself up to a phase, but
acts on the set of degenerate states by
a multi-dimensional unitary transformation.
Let us compute it explicitly: putting all terms together, the
four-point function becomes,
\barr
G(z_i)\times\prod_{i<j}\ z_{ij}^{p+1/2} &= &
\prod_{i<j}\ z_{ij}^{p+1}\ \left[\ A_1\ {\cal F}_1(z_i)\ + \
A_2\ {\cal F}_2(z_i)\ \right]\ , \nl
{\cal F}_1(z_i) &=& \frac{1}{z_{13}z_{24}z_{12}z_{34}}\ ,\qquad
{\cal F}_2(z_i)\ =\ \frac{1}{z_{13}z_{14}z_{24}z_{23}}\ .
\label{elco}
\earr
For example, the exchange of the $(1)$ and $(2)$ electrons,
$(z_1 - z_2) \to {\rm e}^{i\pi}(z_1 - z_2)$ leads to the
following transformation: the prefactor yields the usual minus 
sign for Fermi statistics ($(p+1)$ is odd), while the two blocks 
transform as follows,
\beq
\left(\matrix{ {\cal F}_1 \cr {\cal F}_2}\right)
\ \rightarrow\ 
\left(\matrix{-1 & 1 \cr 0 & 1  }\right)
\left(\matrix{ {\cal F}_1 \cr {\cal F}_2}\right)\ .
\label{naexp}\eeq

The hierarchical Hall states have been commonly considered to posses 
electrons and quasi-particles with Abelian statistics \cite{wen};
this is in fact true if one adopts the
description by the multi-component Abelian CFTs \cite{abe}, where the 
quasi-particles are associated to $\u1$ representations 
and have the usual vertex-operator correlations.
On the other hand, the $(c=2)$ $\winf$ minimal models associate the
quasi-particles to the Virasoro representations, which have assigned an
$SU(2)$ isospin quantum number, leading to non-Abelian 
statistics (and, correspondingly, an $SU(m)$ weight for $\nu=m/(mp\pm 1)$).
These two different descriptions of the quasi-particles
can be tested in future experiments
involving interference and scattering processes, which measure
the three and higher point functions.

Next, we remark that the non-Abelian statistics (\ref{naexp}) occurring
in the $\winf$ minimal models is rather ``mild'', i.e. it is not
due to strong singularities of the correlators; actually, it
has a simple {\it a-posteriori} explanation. 
In the two-component Abelian CFT, there are
two electron fields, one for each component, which are represented
by the vertex operators with $m=1/2$ and $m=-1/2$, respectively.
On the contrary, in the minimal
models there is a single electron excitation represented
by the $s=1/2$ Virasoro primary field: in the Coulomb Gas approach, this
field can be written as either vertex operators with 
$m=\pm 1/2$, due to the reflection symmetry $m \to - m$. 
Moreover, in the electron correlators the charge neutrality can be
satisfied without adding any screening charge: for example,
the two conformal blocks in (\ref{elco}) can also be written as follows:
\barr
{\cal F}_1\ &\propto&\ 
\langle\ \Omega\ \vert\ 
{\bf :}{\rm e}^{-{i \over\sqrt{2}} \varphi(z_1)}{\rm :}
{\bf :}{\rm e}^{{i\over\sqrt{2}} \varphi(z_2)}{\rm :}
{\bf :}{\rm e}^{{i\over\sqrt{2}} \varphi(z_3)}{\rm :}
{\bf :}{\rm e}^{-{i\over\sqrt{2}} \varphi(z_4)}{\rm :}
\ \vert\ \Omega\ \rangle\ , \nl
{\cal F}_2\ &\propto&\ 
\langle\ \Omega\ \vert\  
{\bf :}{\rm e}^{-{i \over\sqrt{2}} \varphi(z_1)}{\rm :}
{\bf :}{\rm e}^{-{i\over\sqrt{2}} \varphi(z_2)}{\rm :}
{\bf :}{\rm e}^{{i\over\sqrt{2}} \varphi(z_3)}{\rm :}
{\bf :}{\rm e}^{{i\over\sqrt{2}} \varphi(z_4)}{\rm :}
\ \vert\ \Omega\ \rangle\ .
\earr
This alternative construction is obtained
by taking all possible sign choices for $m_i=\pm 1/2$ having vanishing sum. 
Therefore, the electrons are represented as in the two-component
Abelian theory, but their components, i.e. signs, are not
distinguishable and should be summed up.
The occurrence of non-Abelian statistics, i.e. of more than one
block, can be traced back to this ambiguity. 
Note that this degeneracy of the electron states is the
type of ``non-Abelian'' statistics also occurring in the 
so-called $[331]$ double-layer Hall state \cite{nonabe}.
Clearly, this fact deserves further investigations before it can be  
physically observed: most notably, one should propose a probe
which couples to the non-trivial neutral excitations.

The non-Abelian statistics has been also investigated in 
the Pfaffian and Haldane-Rezayi $\nu=1/2$ Hall states of 
paired electrons \cite{nonabe}. 
Their quasi-particles are non-Abelian,
but their electrons are still Abelian: actually, for these states 
there is a clear relation between the (bulk) wave-functions 
and the CFT correlators on the edge, which forces
the electrons to obey the usual Fermionic Abelian statistics.
On the contrary, in the $\winf$ models, the electrons are also
non-Abelian ($s=1/2$); actually, the  
relation with wave functions is not immediate: we have compared the 
correlator (\ref{elco}) with the Jain composite-fermion wave 
functions, which have been recently written in the first
Landau level \cite{jaka}. 
We have found that these wave functions do not match
any conformal correlator, because they are not $SL(2,{\bf C})$
covariant and then cannot be written as correlators of Virasoro primary fields. 
Presumably, the electrons are described by different fields 
in the bulk and in the edge. Therefore, the result (\ref{naexp})
for the non-Abelian
statistics of the electron excitations at the edge is consistent
with the experimental and theoretical results known at present. 


\section{Conclusions}

In this work, we have pursued the investigation of the $\winf$
minimal models, which were previously proposed as the conformal theories of
the hierarchical Hall states \cite{ctz5}.
We have explicitly formulated these theories in a bosonic
Fock space with either a set of constraints or a relevant Hamiltonian. 
We have presented a detailed analysis of their physical properties 
and we have compared them with the other, multi-component
Abelian theories \cite{abe}.
We have relied upon several known methods in conformal
field theory, like the Hamiltonian reduction, the
Coulomb Gas approach and the spinon excitations.
This explicit formulation of the $\winf$ minimal models has already 
been useful for interpreting the numerical analysis of 
the low-lying electron spectrum on a disk geometry of Ref.\cite{cmsz}.

Moreover, the Hamiltonian formulation of Section $2$ 
suggests the possibility of further investigations: 
for example, it would be interesting to describe the $\winf$ minimal models 
by means of the Chern-Simons field theory, along the
lines of the well-known relation between edge degrees of freedom
and bulk gauge fields \cite{abe}; this description may reveal 
interesting properties
of the bulk states which support minimal edge excitations.
Another interesting result is the computation of correlation functions
in Section $3$:
this will turn out to be important for searching characteristic signals
of the $\winf$ minimal models in two recently proposed experiments:
the multi-point interferometer of Ref.\cite{multi} and the 
detection of the Andreev reflection \cite{andre}.

\noindent{\bf Acknowledgements}

We would like to thank the C.E.R.N.
Theory Division and the Theory Group at L.A.P.P., Annecy, for hospitality.
A. C. also thanks the Theory Group of the Centro At\'omico 
Bariloche for hospitality and acknowledges the partial
support of the European Community Program FMRX-CT96-0012.
G. R. Z. is grateful to I.N.F.N., Sezione di Firenze,
and I.C.T.P., Trieste, for hospitality; 
his work is supported by a grant of the Antorchas
Foundation (Argentina).

\appendix
\section{Properties of the $c=3$ $\winf$ Minimal Models}

In this Appendix, we describe the $c=3$ $\winf$ minimal
models appropriate for filling fractions $\nu=3/(3p+1)$
($p=2,4,\dots$) following closely the presentation 
of Sections $2$ and $3$. Our aim is to describe the reduction in
the number of degrees of freedom from the $\u1\times\sut$
conformal theories \cite{abe} to the $c=3$ $\winf$ minimal 
models with symmetry algebra $\u1\times {\cal W}_3$ \cite{ctz5}.
We first rewrite the spectrum of quasi-particle charge $Q$ and quantum
statistics $\theta/\pi=2h$ of the $\u1\times\sut$ theories
in a basis which factorizes the $\u1$ part.
Next, we analyze the neutral spectrum according to the \reps
of the nested algebras $\sut \supset \u1^2 \supset \w3$.

The $\u1\times\sut$ spectrum in the basis of Ref. \cite{abe} is:
\barr
Q & =& {n_1+n_2+n_3 \over 3p+1} \ ,\qquad\qquad\qquad\qquad n_1,n_2,n_3 \in 
{\bf Z}\ ,\nl
h & =& {1\over 2}\left( n_1^2 +n_2^2 +n_3^2 \right) -
{p\over 2 (3p+1)}\left( n_1+n_2+n_3 \right)^2 \ .
\label{su3sp}\earr
The integers $n_i$ span the three-dimensional lattice $\Gamma$,
those vectors have norm $h$, the total Virasoro dimension;
each point of the lattice identifies a $\u1^3$ highest-weight \rep 
in this theory. 
A two-dimensional sub-lattice is the $SU(3)$ weight lattice
${\bf P}$; this can be generated by the fundamental weights
$\vec{\Lambda}^{(i)}$, $i=1,2$, which are dual to the simple positive roots
$\vec{\alpha}_{(j)}$. A standard basis is \cite{wyb}:
\barr
{\vec \Lambda} &=& \ell_1\ {\vec \Lambda}^{(1)}\ +\ 
\ell_2\ {\vec \Lambda}^{(2)} \quad \in {\bf P}\ ,\nl
{\vec \Lambda}^{(1)} &=& 
\left( \frac{1}{\sqrt{2}} , \frac{1}{\sqrt{6}}\right)\ , \qquad 
{\vec \Lambda}^{(2)} = \left( 0 , \frac{2}{\sqrt{6}} \right) \ , \nl
{\vec\a}_{(1)} &=& \left( \sqrt{2} , 0 \right)\ , \qquad\qquad
{\vec\a}_{(2)} = \left( - \frac{1}{\sqrt{2}} , \sqrt{{3\over 2}}\right)\ ;
\label{su3latt}\earr
they satisfy
${\vec\a}_{(i)} \cdot {\vec \Lambda}^{(j)} = \delta^j_i$ ,
$\left\Vert {\vec\a}_{(i)} \right\Vert = 2$  and
$\left\Vert {\vec\Lambda}^{(i)} \right\Vert = 2/3$, with $ i,j=1,2$,  

We recall that each $SU(3)$ Lie-algebra \rep can be characterized by
a positive highest weight $\vec{\Lambda} \in {\bf P}^+$, i.e. by the pair
$(\ell_1,\ell_2)$ with $\ell_1,\ell_2 \ge 0$.
Other copies of the \reps have $\ell_i \in {\bf Z}$ and can be obtained
by acting on ${\bf P}^+$ with the Weyl group.
The triality of the $(\ell_1,\ell_2)$ \rep is given by: 
\beq
\a =\ell_1 -\ell_2 \ \ {\rm mod\ } 3 \ ,\quad \a =0, \pm 1\ .
\label{trial}\eeq
We denote with ${\bf P}_\a$ the three sub-lattices of ${\bf P}$ with
given triality (each one is equivalent to the root lattice up to a 
translation).

Form the construction of the lattice $\Gamma$ in Ref. \cite{abe}\cite{ctz5}, 
it is apparent that its factorization into ${\bf P}$ and the one-dimensional
charge lattice can only be done within each $\a$-sector;
moreover, the integers in the two lattices are related by 
$\ell_1=n_1-n_2$ and $\ell_2=n_2-n_3$.
Therefore, we consider the $\a$-dependent change of basis:
\beq
\begin{array}{l l l}
\a =0\ : && \\
3\ell =n_1+n_2+n_3 \ , 
&\ell_1 = n_1 -n_2 = 2k_1 - k_2\ , 
&\ell_2 = n_2 -n_3 = 2k_2 - k_1 \ ;\\
\a =1\ : & & \\
3\ell+1 =n_1+n_2+n_3 \ , 
&\ell_1 = n_1 -n_2 = 2k_1 - k_2+1\ , 
&\ell_2 = n_2 -n_3 = 2k_2 - k_1 \ ;\\
\a =-1\ : &&\\ 
3\ell -1 =n_1+n_2+n_3 \ , 
&\ell_1 = n_1 -n_2 = 2k_1 - k_2\ , 
&\ell_2 = n_2 -n_3 = 2k_2 - k_1 +1\ .
\end{array} 
\label{magtra}\eeq
This rewrites the spectrum (\ref{su3sp}) in the form:
\barr
Q &=& \frac{3 \ell + \a }{3p+1} \ , \nl
h &=& \frac{(3 \ell + \a)^2 }{6(3p+1)}\ +\ h_{\vec \Lambda}\ ,
\label{spsu3}\earr
where $h_{\vec\Lambda}$ is the dimension of the neutral
$\u1^2$ \reps in the lattice ${\bf P}$:
\beq
h_{\vec \Lambda} = {1\over 2} \left\Vert \vec{\Lambda} \right\Vert^2 =
\frac{1}{3}\ \left( \ell^2_1 + \ell^2_2 +\ell_1 \ell_2 \right)\ =
\left\{
\begin{array}{l l}
k^2_1 + k^2_2 - k_1 k_2   \ ,          & \a =0 \\
k^2_1 + k^2_2 - k_1 k_2 + k_1 + {1\over 3} \ ,& \a =\pm 1 \ .
\end{array}
\right.
\label{su3wei}\eeq
The factorization of $\Gamma$ into charged and neutral orthogonal sub-lattices 
is achieved in (\ref{spsu3}): the former is parametrized by
$\ell \in {\bf Z}$; the latter by $\ell_1,\ell_2 \in {\bf Z}$
(lattice ${\bf P}$), or alternatively by $k_1,k_2 \in {\bf Z}$
in each sector ${\bf P}_\a$.

Form now on, we discuss the neutral spectrum only.
We want to analyze the decompositions of its \reps with respect to
the algebras $\sut\supset\u1^2\supset \w3$.
The $\sut$ algebra has three (integrable) \reps labelled
by $\a=0,\pm 1$ \cite{cft}; each of these decomposes:
(i) into the sum of the $\u1^2$ \reps of the lattice ${\bf P}_\a$,
Eqs.(\ref{spsu3}),(\ref{su3dim}), each with multiplicity one;
(ii) into the sum of the $\w3$ \reps of weights 
$\vec{\Lambda} \in {\bf P}^+_\a$, each with
multiplicity given by the dimension of the corresponding $SU(3)$ \rep.
The latter decomposition can be written $\sut\sim SU(3)\times \w3$: 
actually, the $SU(3)$ generators $J^a_0$,
$a=1,\dots, 8$, commute with the $\w3$ ones, $L_n^S$ and ${\bf W}_n$,
which are the modes of the Virasoro and spin-three currents, respectively;
this factorization is completely analogous to
the $\su2$ case of Section $2$.

These decompositions of \reps can be checked by computing the 
corresponding characters. The $\sut$ characters are found in Ref.
\cite{itz}: they involve the sum over the sub-lattice
$M_\a = \left\{ \vec{\Lambda}\vert \vec{\Lambda}=\vec\gamma_o
+ 4 \vec\gamma \right\}$, with 
$\gamma_o=\left\{0,\vec{\Lambda}^{(1)}, \vec{\Lambda}^{(2)}\right\}$,
for $\a=\{0,1,-1\}$, respectively, and $\vec\gamma$ a vector of the
root lattice. Their explicit form is:
\beq
\chi^{\sut}_{\a} = {1 \over {\eta(q)^8}}
\sum_{(a,b)= (i_1,i_2) +4 (2k_1-k_2, 2k_2 - k_1)} 
d(a-1,b-1)\ q^{(a^2 +b^2+a b)/12} \ ,
\label{su3cha}\eeq
with
\barr
(i_1,i_2) &=& \left\{ (1,1), (2,1) , (1,2)\right\}\ ,
\quad {\rm resp.\ for}\ \a =\left\{0,1,-1 \right\}\ , \qquad
{\rm and} \ \ k_1,k_2 \in {\bf Z}; \nl
d_{\vec\Lambda} &\equiv & d\left( \ell_1,\ell_2 \right) =
{1\over 2} ( 1 + \ell_1) (1 + \ell_2) (2 +\ell_1 + \ell_2)\ .
\label{su3dim}\earr
In the last equation, $d_{\vec\Lambda}$ 
is the dimension of the $SU(3)$ \rep $(\ell_1,\ell_2)\in {\bf P}^+$
(here extended to all ${\bf P}$).

In order to verify the first of the decompositions above,
these characters are compared with those obtained by summing
the $\u1^2$ \reps in each sector ${\bf P}_\a$ of the
quasi-particle spectrum (\ref{spsu3}), (\ref{su3wei}).
Using the form (\ref{abecha}) of the $\u1$ character, we
obtain the expressions: 
\barr
\chi^{\sut}_{\a=0} &=& {1\over {\eta(q)^2}} \sum_{k_1,k_2 \in \Z}
q^{k^2_1 + k^2_2 - k_1 k_2}\ , \nl
\chi^{\sut}_{\a=\pm1} &=& {1\over {\eta(q)^2}} \sum_{k_1,k_2 \in \Z}\
q^{k^2_1 + k^2_2 - k_1 k_2 + k_1 + 1/3}\ .
\label{u12cha}\earr
The expressions (\ref{su3cha}) and (\ref{u12cha}) are actually
equivalent, due to non-trivial Jacobi-like identities; these
can be checked by expanding both expressions in
series of $q$ with the help of Mathematica \cite{wolf}.
This completes the analysis of the \rep content of the
$c=3$ Abelian CFT with symmetry $\u1\times\sut$. 

The second, more relevant, decomposition is 
$\sut\sim \w3 \times SU(3)$: we need the $c=2$ $\w3$ characters,
which are obtained from Ref.\cite{kac}. They read:
\beq
\chi^{{\cal W}_3}_{{\bf \Lambda}}\ =\ 
\chi^{{\cal W}_3}_{\ell_1,\ell_2}\ 
=\ {1 \over {\eta(q)^2}}\ 
q^{(\ell^2_1 + \ell^2_2 + \ell_1 \ell_2)/3}\ \left( 1 - q^{\ell_1 +1} \right)\
\left( 1 - q^{\ell_2 +1} \right)\ \left( 1 - q^{\ell_1 +\ell_2 + 2} \right)\ .
\label{wudue}
\eeq
Actually, the $c=3$ degenerate $\winf$ \reps \cite{kac} of
weights $\vec{r}=\{s+n_1,s+n_2,s+n_3\}$, with $s \in {\bf R}$
and $n_1 \ge n_2 \ge n_3 $, are equivalent to the $\u1\times\w3$ \reps
with $\ell_1=n_1-n_2$ and $\ell_2=n_2-n_3$ \cite{ctz5}.
Note that these $\w3$ characters are similar to the degenerate
Virasoro ones (\ref{chiva}), in the sense that 
they are again linear combinations, with alternating signs, of the 
corresponding Abelian characters. In Eq. (\ref{wudue}), 
there are six of them, which sit at the points 
of an hexagon of the root lattice: actually, we can rewrite this
character, e.g. for $\a=0$,
\beq
\chi^{\w3}_{2k_1-k_2,2k_2-k_1} = 
\chi^{\u1^2}_{k_1,k_2} - \chi^{\u1^2}_{k_1+1,k_2}
+ \chi^{\u1^2}_{k_1+2,k_2+1} - \chi^{\u1^2}_{k_1+2,k_2+2}
+ \chi^{\u1^2}_{k_1+1,k_2+2} - \chi^{\u1^2}_{k_1,k_2+1} \ .
\eeq
This describes the decomposition of \reps  for $\u1^2 \supset \w3$ and
shows the existence of $\w3$ null vectors.

According to the equivalence $\sut\sim \w3\times SU(3)$, the
character $\chi^{\sut}_{\a}$ (\ref{su3cha}) should be reproduced by summing
over the $\w3$ ones with weights $\vec{\Lambda}\in {\bf P}_{\a}^+$
and multiplicities equal to the dimensions of the 
corresponding $SU(3)$ multiplets (\ref{su3dim}). These sums are written:
\barr
\chi^{\sut}_{\a =0} &=& \sum_{k_1=0}^{\infty} \ \
\sum_{2k_1 \ge k_2 \ge k_1/2} 
\ d(2k_1 -k_2 , 2k_2 -k_1)\ \chi^{{\cal W}_3}_{2k_1 -k_2 , 2k_2 -k_1}\ , 
\nl
\chi^{\sut}_{\a =1} &=& \sum_{k_1=0}^{\infty}\ \
\sum_{2k_1 +1\ge k_2 \ge k_1/2} 
\ d(2k_1 -k_2 +1, 2k_2 -k_1)\ 
\chi^{{\cal W}_3}_{2k_1 -k_2 +1, 2k_2 -k_1}\ , 
\nl
\chi^{\sut}_{\a =-1} &=& \sum_{k_1=0}^{\infty}\ \
\sum_{2k_1 \ge k_2 \ge (k_1-1)/2} 
\ d(2k_1 -k_2 , 2k_2 -k_1 + 1)\  
\chi^{{\cal W}_3}_{2k_1 -k_2 , 2k_2 -k_1 + 1} \ ,\ \ 
\label{invsu3}\earr
where the ranges for $(k_1,k_2)$ are obtained from $\ell_1,\ell_2\ge 0$. 
The identities (\ref{invsu3}) are again checked by power expansion in $q$. 
Therefore, we have proven the expected decomposition of $\sut$ into
$\w3$: note the complete analogy with the $\su2$ case (\ref{deco})
illustrated in Fig. (\ref{fig1}).

The Hamiltonian reduction from the $\u1\times\sut$ Abelian theory
to the $c=3$ $\winf$ minimal model is then obtained by keeping one state
per $SU(3)$ multiplet; this is the $SU(3)$ lowest-weight state, which
is annihilated by the $SU(3)$ shift operators \cite{wyb}:
\barr
E^{-\vec{\a}_{(1)}}_0\ \vert {\rm minimal\ state} \rangle &=& 0 \ , \nl 
E^{-\vec{\a}_{(2)}}_0\ \vert {\rm minimal\ state} \rangle &=& 0 \ .
\label{su3cst}\earr
These two constraints define the $c=3$ $\winf$ minimal models in the
Abelian Fock space, in analogy with (\ref{minvin}).
Note that the third negative root $\vec{\a}_{(1)}+\vec{\a}_{(2)}$
does not give an independent condition because
$\left[ E^{-\vec{\a}_{(1)}}_0,E^{-\vec{\a}_{(2)}}_0\right] 
= - E^{-\vec{\a}_{(1)}-\vec{\a}_{(2)}}_0$.

The equivalent Hamiltonian formulation of the $c=3$ $\winf$ minimal models is
obtained by adding the following relevant term to the Hamiltonian:
\beq
H^{(2)} = \gamma \left( E^{\vec{\a}_{(1)}}_0\ E^{-\vec{\a}_{(1)}}_0
+ E^{\vec{\a}_{(2)}}_0\ E^{-\vec{\a}_{(2)}}_0 \right) \ ,
\eeq
and by considering the infrared limit $\gamma\to\infty$ of the corresponding
renormalization-group trajectory.
 
Finally, we present the Fermionic realization of the $c=3$ $\winf$ minimal
models that parallels that of Section $3$. Consider three Weyl chiral 
fields $\Psi_i$, $i=1,2,3$. We can realize the $\u1\times\sut$ theory
in terms of fermion bilinears ${\bf :}{\overline{\Psi}}_i O^{ij}
\Psi_j{\bf :}$, where the matrices $O^{ij}$ represent the fundamental of
$SU(3)$. The Cartan sub-algebra contain the charge current,
\beq
J\ =\left( {\overline{\Psi}}_1 \Psi_1 +{\overline{\Psi}}_2 \Psi_2 +
{\overline{\Psi}}_3 \Psi_3 \right)\ ,
\eeq
and the neutral currents,
\beq
H_1\ ={1\over \sqrt{2}}
\left( {\overline{\Psi}}_1 \Psi_1 - {\overline{\Psi}}_2 \Psi_2  \right)\ ,
\qquad
H_2\ ={1\over \sqrt{6}} \left(
{\overline{\Psi}}_1 \Psi_1 +{\overline{\Psi}}_2 \Psi_2 -2\ 
{\overline{\Psi}}_3 \Psi_3 \right)\ .
\eeq
The shift currents are:
\beq
E^{\vec{\a}_{(1)}} = {\overline{\Psi}}_1 \Psi_2 \ ,\qquad 
E^{\vec{\a}_{(2)}}= {\overline{\Psi}}_2 \Psi_3 \ ,\qquad 
E^{\vec{\a}_{(1)}+\vec{\a}_{(2)}} = {\overline{\Psi}}_1 \Psi_2 \ ,
\label{curasu}\eeq
together with their Hermitian conjugates.
By construction, the zero-modes of these currents satisfy the standard 
commutation relations of the $SU(3)$ Lie algebra in the Cartan basis
\cite{wyb}.
As in Section $3$, the constraints (\ref{su3cst}) can be 
written in the three-component Fermionic Fock space, and they imply that 
the states of the minimal-model are made by particle-hole excitations
which are completely symmetric with respect to the three components.
The constraints can be interpreted as further incompressibility conditions
which eliminate the anti-symmetric excitations tangential to the Fermi 
surface.

\def\NP{{\it Nucl. Phys.\ }}
\def\PRL{{\it Phys. Rev. Lett.\ }}
\def\PL{{\it Phys. Lett.\ }}
\def\PR{{\it Phys. Rev.\ }}
\def\CMP{{\it Comm. Math. Phys.\ }}
\def\IJMP{{\it Int. J. Mod. Phys.\ }}
\def\MPL{{\it Mod. Phys. Lett.\ }}
\def\RMP{{\it Rev. Mod. Phys.\ }}
\def\AP{{\it Ann. Phys.\ }}

\end{document}